\documentclass[final,nomarks,nohyperref]{dmtcs-episciences}
\newif\ifDRAFT
\DRAFTfalse

\usepackage[all]{xy}
\usepackage{bussproofs}
\usepackage{listings}
\usepackage{alltt}
\usepackage{qsymbols}
\usepackage[utf8]{inputenc}
\usepackage{amssymb}
\usepackage{stmaryrd}
\usepackage{tikzsymbols}
\usepackage{MnSymbol}
\usepackage{wasysym}
\ifDRAFT \usepackage{draftwatermark} \SetWatermarkScale{3.5}  \SetWatermarkLightness{0.85}\fi

\usepackage[shortcuts]{extdash}

\newif\ifMint
\Mintfalse
\ifMint \usepackage{minted}  \usepackage{ifsym} \usepackage{csquotes}

\else
\fi
\definecolor{darkGreen}{rgb}{0,.5,0}
\definecolor{mauve}{rgb}{1,0,1}
\definecolor{rougefonce}{cmyk}{.3,1,.3,0}
\definecolor{cyanp}{cmyk}{.5,.3,0,0}
\definecolor{yellow}{cmyk}{0,0,.7,0}
\definecolor{beige}{cmyk}{0,.2,.7,0}
\definecolor{brun}{cmyk}{0,.5,.7,0}
\definecolor{darkBrown}{cmyk}{.3,.75,.75,.15}
\newcommand{\rouge}[1]{{\color{red} #1}}

\newcommand{\bl}[1]{\textcolor{blue}{#1}}

\newcommand{\brunir}[1]{{\color{darkBrown} #1}}

  {\color{blue}}
  {}
  {\color{darkBrown}}
  {}

  {\color{red}}
  {} %

\newcommand{\node}[1]{*++[o][F]{\scriptstyle #1}}
\newcommand{\snode}{*+<2pt>[o][F]{\scriptstyle ~}}
\newcommand{\msnode}{*+<2pt>[o][F]{\scriptstyle \star}}
\newcommand{\carre}[1]{*++[F]{\scriptstyle #1}}
\newcommand{\scarre}{*+<4pt>[F]{\scriptstyle \phantom{.}}}
\newcommand{\CARRE}[1]{*++[F=]{\scriptstyle #1}}
\newcommand{\sCARRE}{*+<2pt>[F]{\scriptstyle \star}}
\newcommand{\sTRIANGLE}{*+<-8pt>[]{\raisebox{-12pt}{$\blacktriangle$}}}
\newcommand{\noeud}[1]{*++[o][F]{\scriptstyle #1} \ar@{-}[dl]\ar@{-}[dr]}
\newcommand{\nfl}{*++[o][F]\to \ar@{-}[dl]\ar@{-}[dr]}
\newcommand{\lf}[1]{*++[F]{\scriptstyle #1}}
\newcommand{\baisse}[2]{\raisebox{-#1}{\ensuremath{#2}}}

\newcommand{\one}{$\Circle\hspace*{-11pt}\nearrow$}
\newcommand{\onw}{$\Circle\hspace*{-11pt}\nwarrow$}
\newcommand{\rla}{\stackrel{\leftarrow}{\rightarrow}}
\newcommand{\dua}{\downarrow\uparrow}

\newcommand{\Keywords}

  \title[Efficient random generation of binary trees]{Holonomic equations and efficient random generation of binary trees}
   \author{Pierre Lescanne}
   \affiliation{LIP, \'Ecole Normale Sup\'erieure de Lyon, France}
   \publicationdata{25:2}{2023}{10}{10952}{2023-02-13}{2023-07-07}
\begin{document}
  
\maketitle

\begin{abstract}
  Holonomic equations are recursive equations which allow computing
  efficiently numbers of combinatoric objects.  Rémy showed that the
  holonomic equation associated with binary trees yields an efficient
  linear random generator of binary trees.  I extend this paradigm to
  Motzkin trees and Schröder trees and show that despite slight
  differences my algorithm that generates random Schröder trees has linear
  expected complexity and my algorithm that generates Motzkin trees is in
  $O(n)$ expected complexity, only if we can implement a specific oracle
  with a $O(1)$ complexity.  For Motzkin trees, I propose a solution which
  works well for realistic values (up to size ten millions)  and yields an
  efficient algorithm.
  \end{abstract}

\section{Introduction}
\label{sec:intro}

In this paper, I am interested in efficient algorithms for randomly
generating several sorts of binary trees. For this, I consider recurrences
defining sequences of integer coefficients and, more precisely, I am
interested in specific recurrences $F_n$ called ``holonomic recurrence''
where roughly speaking, ``holonomic'' means that $F_{n+s}$ is a
combination, using polynomials in $n$, of the $F_i$'s, for
${n\le i\le n+s}$.  More precisely, (see Flajolet and Sedgewick's
book~\cite{flajolet08:_analy_combin}, Appendix B.4) the coefficients
fulfill the following recurrence:
\begin{displaymath}
  P_{s}(n) F_{n+s} + P_{s-1}(n) F_{n+s-1 } + ... + P_0(n) F_{n} = 0
\end{displaymath}
for some $n\ge n_0$, where the $P_j(n)$ are polynomials in $n$.
This kind of recurrence is called a \emph{$P$-recurrence}.  For instance, 
for Catalan numbers:
\begin{displaymath}
  C_n -  \sum_{k=0}^{n-1} C_k C_{n-k-1} = 0
\end{displaymath}
is the classical recurrence that is used in general to defined them, but it is not a $P$-recurrence, whereas 
\begin{displaymath}
  (n+1)  C_{n}- 2(2n-1) C_{n-1} = 0
\end{displaymath}
is the $P$-recurrence, which will be considered later on.  Notice that
initial values should be added to this $P$-recurrence. This will be
considered in the paper for each specific case.

In this paper, I consider three families of binary trees (planar binary
trees or Catalan trees,
Motzkin trees aka unary-binary trees, Schröder trees) and their random
generation. It turns out that holonomic recurrences play a key role in the
design of efficient random generation algorithms.

The three examples: binary trees, Motzkin trees, Schröder trees are
interesting because they have different holonomic equations, one (Catalan
numbers) has one term on the right, one (Motzkin numbers) has a sum of two
terms on the right and one (Schröder numbers) has a subtraction of two
terms, on the right.  These yield different random generation algorithms,
as this will be explained further in this paper.

This paper is associated with a
\href{https://github.com/PierreLescanne/Motzkin}{library} of programs
written in \textsf{Haskell} and in \textsf{Python}, available on GitHub.
\ifMint \else The reader who wants to read a better
rendering of the programs of this paper is invited to get the version on
my \href{http://perso.ens-lyon.fr/pierre.lescanne/PUBLICATIONS/Hol_Lin_Gen.pdf}{web
  page} or the  \href{https://github.com/PierreLescanne/Motzkin}{version available on GitHub}.\fi

\section{Random binary trees}
\label{sec:RandBinTree}

\begin{figure}[!]
  \centering
   \begin{math}
    \raisebox{3pt}{
    \xymatrix@C=6pt @R=6pt{
      &&\ar@{.}[d]\\
      &&*=<3pt>{\scriptstyle \ostar}\ar@{-}[ddll]\ar@{-}[ddrr]\\
      &&\mathsf{T}\\
      *=<1pt>{}\ar@{-}[rrrr]&&&&*=<1pt>{} }}
    \qquad \xymatrix@C=6pt @R=6pt{
      &&&\ar@{.}[d]\\
      &&&\snode\ar@{-}[dl]\ar@{-}[dr]\\
      &&*=<3pt>{\circ}\ar@{-}[ddll]\ar@{-}[ddrr]&&\sCARRE\\
      &&\mathsf{T}\\
      *=<1pt>{}\ar@{-}[rrrr]&&&&*=<1pt>{} }
  \end{math}
  \caption{Rémy's right insertion of a leaf}
  \label{fig:Remy}
\end{figure}
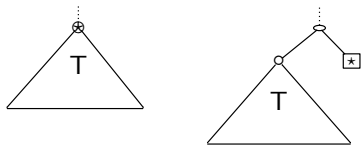

\emph{Rémy's algorithm}~\cite{DBLP:journals/ita/Remy85} for generation of
random binary trees is of linear complexity, i.e., $O(n)$.  It is based on a constructive proof of the holonomic equation~\cite{rodrigues38}:
\begin{displaymath}
    (n+1)  C_{n}= 2(2n-1) C_{n-1}
  \end{displaymath}
  Here ``constructive'' means that an explicit bijection between objects
  counted by the both sides of the equation is provided.  In the case of
  binary trees, this holonomic equation is very peculiar since $C_n$ times
  a polynomial in $n$ is equal to $C_{n-1}$ times a polynomial in $n$.  We
  will see that this is not the case for Motzkin numbers and Schröder
  numbers, but the paradigm can be extended.  Rémy's algorithm is
  described by Knuth in~\cite{KnuthVol4_4} § 7.2.1.6 (pp.~18-19) and works
  on \emph{extended binary trees}, or just \emph{binary trees} in which we
  distinguish \emph{internal nodes} and \emph{external nodes} or
  \emph{leaves}.  The idea of the algorithm is that a random binary tree
  can be built by iteratively and randomly drawing an internal node or a
  leaf in a random binary tree and inserting, between it and its parent a
  new internal node and a new leaf either on the left or on the right (see
  Figure~\ref{fig:Remy}).  An insertion is also possible at the root.  In
  this case, the new inserted node becomes the root.  The root can be seen
  as the child of a hypothetical node

A binary tree of size $n$ has $n-1$ internal nodes and $n$ leaves.  We
label binary trees with numbers between $0$ and $2n-2$ such that internal
nodes are labeled with odd numbers and leaves are labeled with even
numbers.  Inserting a node in a binary tree of size $n$ requires drawing
randomly a number between $0$ and $4n-3$.  This process can be optimized
by representing a binary tree as a list (a \texttt{vector} in \textsf{Haskell}), an
idea sketched by Rémy and described by Knuth.  In this vector, even values
are for internal nodes and odd values are for leaves.  The root is located
at index $0$. The left child of an internal node with label $2k+1$ is
located at index $2k+1$ and its right child is located at index $2k+2$.
Here is a vector representing a binary tree with 10 leaves and its
drawing.

\begin{displaymath}
  \begin{array}{|l||r|r|r|r|r|r|r|r|r|r|r|r|r|r|r|r|r|r|r|r|r|r|}
    \hline
    \textbf{\footnotesize indices}&0&1&2&3&4&5&6&7&8&9&10&11&12&13&14&15&16&17&18\\
    \hline
    \textbf{\footnotesize values}&1&13&0&2&5&9&7&8&4&11&17&12&10&15&3&16&14&18&6\\
    \hline
  \end{array}
\end{displaymath}
\begin{displaymath}
  \xymatrix @C=5pt @R=5pt{
    &&&\node{1}\ar@{-}[dl]\ar@{-}[dr]\\ 
    &&\node{13}\ar@{-}[dl]\ar@{-}[drr]&&\lf{0}\\ 
    &\node{15}\ar@{-}[dl]\ar@{-}[dr]&&&\node{3}\ar@{-}[dl]\ar@{-}[dr]\\ 
    \lf{16}&&\lf{14}&\lf{2}&&\node{5}\ar@{-}[dl]\ar@{-}[drr]\\ 
    &&&&\node{9}\ar@{-}[dll]\ar@{-}[dr]&&&\node{7} \ar@{-}[dl]\ar@{-}[dr]&\\ 
    &&\node{11}\ar@{-}[dl]\ar@{-}[dr] &&& \node{17}\ar@{-}[dl]\ar@{-}[dr] &\lf{8}&&\lf{4}\\ 
    &\lf{12 }&& \lf{10} & \lf{18}&&\lf{6}
  }
\end{displaymath}

This tree was built by inserting the node $17$ together with the leaf $18$
in the following vector.
\begin{displaymath}
  \begin{array}{|l||r|r|r|r|r|r|r|r|r|r|r|r|r|r|r|r|r|r|r|r|}
    \hline
    \textbf{\footnotesize indices}&0&1&2&3&4&5&6&7&8&9&10&11&12&13&14&15&16\\
    \hline
    \textbf{\footnotesize values}&1&13&0&2&5&9&7&8&4&11&6&12&10&15&3&16&14\\
    \hline
  \end{array}
\end{displaymath}
which codes the tree
\begin{displaymath}
  \xymatrix @C=5pt @R=5pt{
    &&&\node{1}\ar@{-}[dl]\ar@{-}[dr]\\ 
    &&\node{13}\ar@{-}[dl]\ar@{-}[drr]&&\lf{0}\\ 
    &\node{15}\ar@{-}[dl]\ar@{-}[dr]&&&\node{3}\ar@{-}[dl]\ar@{-}[dr]\\ 
    \lf{16}&&\lf{14}&\lf{2}&&\node{5}\ar@{-}[dl]\ar@{-}[drr]\\ 
    &&&&\node{9}\ar@{-}[dll]\ar@{-}[dr]&&&\node{7} \ar@{-}[dl]\ar@{-}[dr]&\\ 
    &&\node{11}\ar@{-}[dl]\ar@{-}[dr] &&& \CARRE{\scriptstyle \brunir{\textbf{\underline{6}}}} &\lf{8}&&\lf{4}\\ 
    &\lf{12 }&& \lf{10}
  }
\end{displaymath}
This was done by drawing a node (internal node or leaf, here the node with
label $6$, right child of the node with label $9$) and a direction (here
right) and by inserting above this node a new internal node (labeled $17$)
and, below the new inserted internal node, a new leaf of the left (labeled
$18$).  This double action (inserting the internal node and attaching the
leaf) is done by choosing a number in the interval $[0..33]$ (in general,
in the interval $[0..(4n-3)]$). Assume that in this case the random
generator returns~$21$.  $21$ contains two pieces of information : its
parity (a boolean) and floor of its half.  Half of $21$ is $10$, which
tells that the new node $17$ must be inserted above the $11^{th}$ node (in
the vector) namely~$6$. Since $21$ is odd, the rest of the tree (here
reduced to the leaf $6$) is inserted on the right (otherwise it would be
inserted on the left).  A new leaf $18$ is inserted on the left (otherwise
it would be inserted on the right).

Consider the same tree and suppose that the random value is $8$. Half of
$8$ is $4$. Hence the new internal node labeled by $17$ is inserted above the node labeled by
$5$
\begin{displaymath}
  \xymatrix @C=5pt @R=5pt{
    &&&\node{1}\ar@{-}[dl]\ar@{-}[dr]\\ 
    &&\node{13}\ar@{-}[dl]\ar@{-}[drr]&&\lf{0}\\ 
    &\node{15}\ar@{-}[dl]\ar@{-}[dr]&&&\node{3}\ar@{-}[dl]\ar@{-}[dr]\\ 
    \lf{16}&&\lf{14}&\lf{2}&&*++[o][F=]{\scriptstyle \brunir{\textbf{\underline{5}}}}\ar@{-}[dl]\ar@{-}[drr]\\ 
    &&&&\node{9}\ar@{-}[dll]\ar@{-}[dr]&&&\node{7} \ar@{-}[dl]\ar@{-}[dr]&\\ 
    &&\node{11}\ar@{-}[dl]\ar@{-}[dr] &&& \lf{6} &\lf{8}&&\lf{4}\\ 
    &\lf{12 }&& \lf{10}
  }
\end{displaymath}
and, since $8$ is even, the rest of the tree is inserted on the left and a
new leaf (labeled $18$) is inserted on the right.
\begin{displaymath}
  \xymatrix @C=5pt @R=5pt{
    &&&\node{1}\ar@{-}[dl]\ar@{-}[dr]\\ 
    &&\node{13}\ar@{-}[dl]\ar@{-}[drr]&&\lf{0}\\ 
    &\node{15}\ar@{-}[dl]\ar@{-}[dr]&&&\node{3}\ar@{-}[dl]\ar@{-}[dr]\\ 
    \lf{16}&&\lf{14}&\lf{2}&&\node{17}\ar@{-}[dl]\ar@{-}[dr]\\
    &&&&\node{5}\ar@{-}[dl]\ar@{-}[drr]&&\lf{18}\\ 
    &&&\node{9}\ar@{-}[dll]\ar@{-}[dr]&&&\node{7} \ar@{-}[dl]\ar@{-}[dr]&\\ 
    &\node{11}\ar@{-}[dl]\ar@{-}[dr] &&& \lf{6} &\lf{8}&&\lf{4}\\ 
    \lf{12 }&& \lf{10}
  }
\end{displaymath}
The algorithm (Figure~\ref{fig:programRemy}) works as follows. If $n=0$,
Rémy's algorithm returns the vector starting at~$0$ and filled with
anything, since the whole algorithm works on the same vector with the same
size.  In general, say that, for $n-1$, Rémy's algorithm returns a vector
$v$ (\texttt{vector} is the concept used in \textsf{Haskell} for arrays
that can be changed in place).  In our \textsf{Haskell} implementation the
function yields an object of type \texttt{Gen~(Vector~Int)} which returns
vector and carries a hidden random number generator.  One accesses to the
generator by \textsf{get} and stores the new generator by \textsf{put}.
One draws a random integer $x$ between $0$ and $4n-3$. Let $k$ be half of
$x$.  In the vector $v$ one replaces the $k^{th}$ position with $2n-1$ and
one appends two elements, namely the $k^{th}$ item of $v$ followed by $2n$
if $x$ is even and $2n$ followed by the $k^{th}$ item of $v$ if $x$ is
odd.

The algorithm builds a uniformly random \emph{decorated binary tree},
i.e., a binary tree with its leaves numbered $0$, $2$,... $2n$.  We notice
that the construction of a tree with such labels is unique, the labels of
the internal nodes are a consequence of the construction, hence are
deduced from the labels of the leaves. If we ignore the leaves, we get a
uniform distribution for the undecorated binary trees (i.e., with no
labels on the leaves).

In the program, \textsf{rands} is a vector of random floating numbers
between $0$ and $1$. 
\begin{figure}[!]
  \centering \ifMint
\begin{minted}{haskell}
rbt :: Int -> Int -> Gen (Vector Int)
rbt seed 0 = do put (mkStdGen seed)
                return(initialVector // [(0,0)])
rbt seed n =
  do v <- rbt seed (n-1)
     generator <- get
     let (rand, newGenerator) = randomR (0::Double,1) generator
     put newGenerator
     let x = floor (rand * fromIntegral (4*n-3))
          -- x is a random value between 0 and 4n-3 --
         k = x `div` 2
     case even x of
       True -> return(v // [(k,2*n-1),(2*n-1,v!k),(2*n,2*n)])
       False -> return(v // [(k,2*n-1),(2*n-1,2*n),(2*n,v!k)])
\end{minted}
 \else
\begin{lstlisting}
rbt :: Int -> Int -> Gen (Vector Int)
rbt seed 0 = do put (mkStdGen seed)
                return(initialVector // [(0,0)])
rbt seed n =
  do v <- rbt seed (n-1)
     generator <- get
     let (rand, newGenerator) = randomR (0::Double,1) generator
     put newGenerator
     let x = floor (rand * fromIntegral (4*n-3))
          -- x is a random value between 0 and 4n-3 --
         k = x `div` 2
     case even x of
       True -> return(v // [(k,2*n-1),(2*n-1,v!k),(2*n,2*n)])
       False -> return(v // [(k,2*n-1),(2*n-1,2*n),(2*n,v!k)])
\end{lstlisting}
  \fi
  \caption{Haskell program for Rémy's algorithm}
  \label{fig:programRemy}
\end{figure}

\section{Motzkin trees}
\label{sec:M-trees}

\emph{Motzkin trees} are also called \emph{unary-binary trees}. This paper
proposes an algorithm for random generation of Motzkin trees.  The
algorithm takes the same paradigm as this of Rémy's linear algorithm for
random generation of \emph{binary trees}~\cite{DBLP:journals/ita/Remy85}.
Assume $n$ is the size of the trees.
  My algorithm for random
  generation of Motzkin trees is based on a bijective proof due to Dulucq
  and Penaud~\cite{DBLP:journals/dm/DulucqP02} of the inductive equality:
  \begin{displaymath}
    (n+2) M_n = (2n+1) M_{n-1} + 3 (n-1)M_{n-2}
  \end{displaymath}
  where the $M_n$'s are the Motzkin numbers.  At some point of the
  algorithm, an ``oracle'' choices which subprogram to call, based on
  $M_{n-1}$ and $M_{n-2}$.  Since $M_{n-1}$ and $M_{n-2}$ are big numbers, this
  induces potentially a not $O(1)$ computation.  A preprocessing allows a
  constant time computation for the oracle. 

  \section{Motzkin numbers and Motzkin trees}
  \label{sec:MotzkinTrees}
  The $n^{th}$ Motzkin number $M_n$ is the number of different ways of
  drawing non-intersecting chords between $n$ points on a circle (not
  necessarily touching every point by a chord).  Motzkin numbers count
  also well parenthesized expressions with a constant $\textbf{c}$, called
  \emph{Motzkin words}. They are words of length $n$ in the language
  generated by the grammar $M$.
  \begin{displaymath}
    M ~=~ \varepsilon \mid \mathbf{c}\,M  \mid \mathbf{(} M \mathbf{)}\,M
  \end{displaymath}

\begin{figure}[!]
  \centering
  \begin{tabular}[c]{c@{\qquad}c@{\qquad}c}
    \includegraphics[width=.1\textwidth]{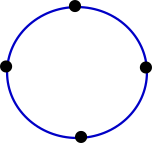}& %
    \includegraphics[width=.1\textwidth]{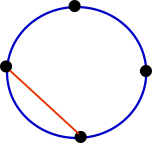}& %
 \includegraphics[width=.1\textwidth]{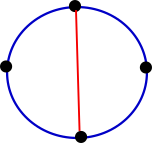}\\\\ %
    \includegraphics[width=.1\textwidth]{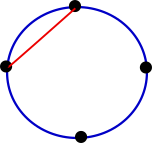}& %
 \includegraphics[width=.1\textwidth]{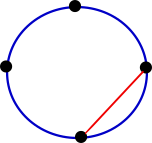}& %
 \includegraphics[width=.1\textwidth]{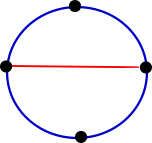}\\\\
    \includegraphics[width=.1\textwidth]{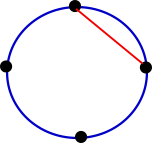}&
 \includegraphics[width=.1\textwidth]{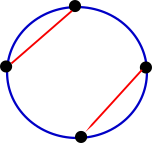}&
 \includegraphics[width=.1\textwidth]{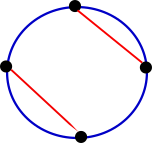}
  \end{tabular}
  \caption[Non-intersecting chords (4 points)]{The 9 non-intersecting
    chords between 4 points on a circle}
  \label{fig:chords4}
\end{figure}

\begin{figure}[!]
  \centering
  \begin{displaymath}
    \begin{array}[c]{c@{\qquad}c@{\qquad}c}
      \mathbf{c}\mathbf{c}\mathbf{c}\mathbf{c}&
                                                \mathbf{c}\mathbf{c}()&\mathbf{c}(\mathbf{c})\\
      \mathbf{c}()\mathbf{c}&(\mathbf{c}\mathbf{c})&(\mathbf{c})\mathbf{c}\\
      ()\mathbf{c}\mathbf{c}&(())&()()
    \end{array}
  \end{displaymath}
  \caption[Parenthesized words]{The 9 parenthesized words with constant
    \textbf{c}.}
  \label{fig:words}
\end{figure}

The bijection between sets of non intersecting chords and well
parenthesized words with constant \textbf{c} is as follows: first one
numbers nodes on the circle counterclockwise, as follows:
\begin{displaymath}
  \includegraphics[width=.15\textwidth]{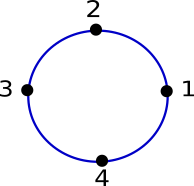}
\end{displaymath}
A position at the beginning of a chord on the circle corresponds to an
opening parenthesis. A position at the end of a chord on the circle
corresponds to a closing parenthesis. A position which is neither of those
corresponds to the constant \textbf{c}.

Motzkin numbers count also routes in the upper quadrant from $(0,0)$ to
$(0,4)$ with move \emph{up}, \emph{down} and \emph{straight}.
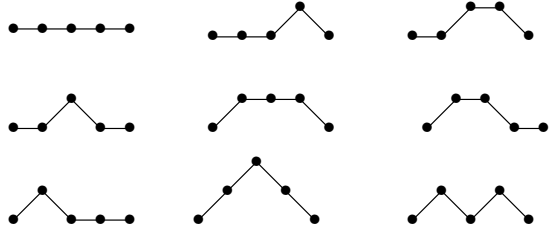
\begin{figure}[!]
  \centering \newcommand{\point}{*=0{\bullet}}
  \begin{tabular}[c]{c@{\qquad}c@{\qquad}c}
    \xymatrix @C = 5pt @R = 5pt{
    &&&&\\
    &&&&\\
    \point \ar@{-}[r]&\point \ar@{-}[r]&\point \ar@{-}[r]&\point \ar@{-}[r]&\point
                                                                             }
                     & %
                       \xymatrix @C = 5pt @R = 5pt{
                                       &&&&\\
    &&&\point \ar@{-}[dr]&\\
    \point \ar@{-}[r]&\point \ar@{-}[r]&\point \ar@{-}[ur]&&\point}
      & %
        \xymatrix @C = 5pt @R = 5pt{
                     &&&&\\
    &&\point\ar@{-}[r]&\point\ar@{-}[rd]&\\
    \point\ar@{-}[r]&\point\ar@{-}[ur]&&&\point&
                                                 }
    \\
    \xymatrix @C = 5pt @R = 5pt{
    &&&&\\
    &&\point\ar@{-}[dr]&\\
    \point\ar@{-}[r]&\point\ar@{-}[ur]&&\point\ar@{-}[r]&\point
                                                          }
      &
        \xymatrix @C = 5pt @R = 5pt{
                     &&&&\\
    &\point\ar@{-}[r]&\point\ar@{-}[r]&\point\ar@{-}[dr]\\
    \point\ar@{-}[ur]&&&&\point
                          }
    &
      \xymatrix @C = 5pt @R = 5pt{
      &&&&\\
    &\point\ar@{-}[r]&\point\ar@{-}[dr]&\\
    \point\ar@{-}[ur]&&&\point\ar@{-}[r]&\point
                                          }
    \\
    \xymatrix @C = 5pt @R = 5pt{
    &&&&\\
    &\point\ar@{-}[dr]&&&\\
    \point\ar@{-}[ur]&&\point\ar@{-}[r]&\point\ar@{-}[r]&\point
                                                          }
      &
        \xymatrix @C = 5pt @R = 5pt{
                     &&\point\ar@{-}[dr]&&&\\
    &\point\ar@{-}[ur]&&\point\ar@{-}[dr]& \\
    \point\ar@{-}[ur]&&&&\point
                          }
    &
      \xymatrix @C = 5pt @R = 5pt{
      &&&&\\
    &\point\ar@{-}[dr]&&\point\ar@{-}[dr]&&\\
    \point\ar@{-}[ur]&&\point\ar@{-}[ur]&&\point
                                           }                         
  \end{tabular}
  \caption[Motzkin paths]{The 9 routes in the upper right quadrant from
    $(0,0)$ to $(0,4)$ with move \emph{up, down and straight}}
  \label{fig:paths}
\end{figure}

The bijection is as follows: an opening parenthesis corresponds to an
\emph{up}, a closing parenthesis corresponds to a \emph{down} and the
constant \textbf{c} corresponds to a \emph{straight}.

Motzkin number $M_n$ counts also the number in unary-binary planar trees
with $n$ edges, that are tree structures with nodes of arity one or two
and with $n$ edges.  Let us call this number $n$ of edges the \emph{size}
of the Motzkin tree.  Notice that the number of nodes of a Motzkin tee of
size $n$ is $n+1$, i.e., the size plus one.  Figure~\ref{fig:truc} gives
the trees for $n=4$.  The bijection $f$ from well parenthesized
expressions with constant \textbf{c} to Motzkin trees is as follows. Its
inverse $f^{-1}$ is also given.
\begin{displaymath}
  \begin{array}{lcl}
    f[\varepsilon] &=& \bullet \\[3pt]
    f[\mathbf{c}~w] &=& \raisebox{10pt}{$\xymatrix@R =8pt@C=5pt{*=0{\bullet}\ar@{-}[d]\\f[w]}$}\\[3pt]
    f[(w_1)~w_2] &=& \raisebox{10pt}{$\xymatrix@R =5pt@C=5pt{&*=0{\bullet}\ar@{-}[dl]\ar@{-}[dr]\\f[w_1]&&f[w_2]}$}
  \end{array}
  \qquad\qquad\qquad
  \begin{array}{lcl}
    f^{-1}[\bullet] &=& \varepsilon \\[3pt]
    f^{-1}\left[\raisebox{5pt}{$\xymatrix@R =8pt@C=5pt{*=0{\bullet}\ar@{-}[d]\\t}$}\right] &=& \mathbf{c}~f^{-1}[t]\\[12pt]
    f^{-1}\left[\raisebox{5pt}{$\xymatrix@R =5pt@C=5pt{&*=0{\bullet}\ar@{-}[dl]\ar@{-}[dr]\\t_1&&t_2}$}\right] &=& (f^{-1}[t_1]) ~ f^{-1}[t_2]
  \end{array}
\end{displaymath}

Motzkin numbers fulfill the equation:
\begin{displaymath}
  M_{n+1} = M_n +\sum_{i=0}^{n-1} M_i\,M_{n-1-i}
\end{displaymath}

\section{Dulucq-Penaud bijection proof}
\label{sec:Dulucq}

As I said, Motzkin numbers fulfill the \emph{holonomic equation}~\cite{flajolet08:_analy_combin}:
\begin{displaymath}
  (n+2) M_n = (2n+1) M_{n-1} + 3 (n-1)M_{n-2}.
\end{displaymath}
Together with the equalities $M_0=1$ and $M_1=1$, Motzkin numbers can be
computed and form the sequence \href{https://oeis.org/A001006}{A001006} in
the \emph{online encyclopedia of integer sequences}~\cite{OEIS}.  In this
section, I~present Dulucq and Penaud's proof of this
equality~\cite{DBLP:journals/dm/DulucqP02}.  This proof relies on the
exhibition of a bijection between the objects counted by the left-hand
side and those counted by the right-hand side.  The first idea is to
consider specific binary trees called \emph{slanting binary trees} and
divide those trees into $7$ subclasses.

\subsection{Slanting binary trees}
\label{sec:slanting}
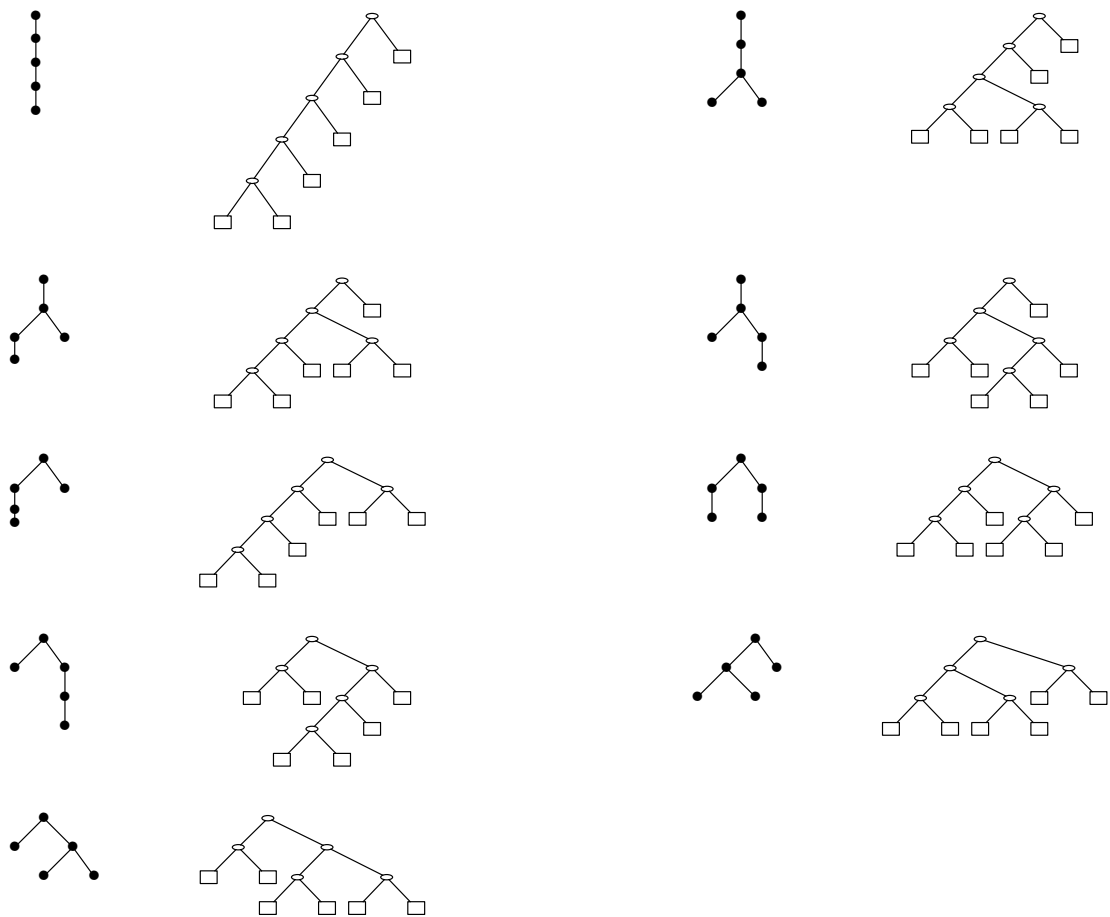
\begin{figure}[!]
  \centering
  \begin{displaymath}
    \begin{array}[c]{c@{\qquad\qquad}c@{\hspace*{100pt}}c@{\qquad\qquad}c}
      \xymatrix @C = 5pt @R = 9pt{
      *=0{\bullet}\ar@{-}[d]\\
      *=0{\bullet}\ar@{-}[d]\\
      *=0{\bullet}\ar@{-}[d]\\
      *=0{\bullet}\ar@{-}[d]\\
      *=0{\bullet}
      }&
         \xymatrix @C = 5pt @R = 9pt{
      &&&&&\snode\ar@{-}[ld]\ar@{-}[rd]\\
       &&&&\snode\ar@{-}[ld]\ar@{-}[rd]&&\scarre\\
       &&&\snode\ar@{-}[ld]\ar@{-}[rd]&&\scarre\\
       &&\snode\ar@{-}[ld]\ar@{-}[rd]&&\scarre\\
       &   \snode\ar@{-}[ld]\ar@{-}[rd]&&\scarre\\
      \scarre&&\scarre
                }    
      & 
        \xymatrix @C = 5pt @R = 5pt{
        &*=0{\bullet}\ar@{-}[d]\\
       &*=0{\bullet}\ar@{-}[d]\\
       &*=0{\bullet}\ar@{-}[ld]\ar@{-}[rd]\\
      *=0{\bullet}&&*=0{\bullet}
                     }&
                        \xymatrix @C = 5pt @R = 5pt{
        &&&&\snode\ar@{-}[ld]\ar@{-}[rd]\\
       &&&\snode\ar@{-}[ld]\ar@{-}[rd]&&\scarre\\
       &&\snode\ar@{-}[ld]\ar@{-}[rrd]&&\scarre\\
       &\snode\ar@{-}[ld]\ar@{-}[rd]&&&\snode\ar@{-}[ld]\ar@{-}[rd]\\
      \scarre&&\scarre&    \scarre&&\scarre
                                     }
      \\\\
      \xymatrix @C = 5pt @R = 5pt{
       &*=0{\bullet}\ar@{-}[d]\\
       &*=0{\bullet}\ar@{-}[dl]\ar@{-}[dr]\\
      *=0{\bullet}\ar@{-}[d]&&*=0{\bullet}\\
      *=0{\bullet}
      }
       &
         \xymatrix @C = 5pt @R = 5pt{
      &&&&\snode\ar@{-}[ld]\ar@{-}[dr]\\
       &&&\snode\ar@{-}[dl]\ar@{-}[drr]&&\scarre\\
       &&\snode\ar@{-}[dl]\ar@{-}[dr]&&&\snode\ar@{-}[dl]\ar@{-}[dr]\\
       &\snode\ar@{-}[dl]\ar@{-}[dr]&&\scarre&\scarre&&\scarre\\
      \scarre&&\scarre
                }     
      & 
        \xymatrix @C = 5pt @R = 5pt{
        &*=0{\bullet}\ar@{-}[d]\\
       &*=0{\bullet}\ar@{-}[dl]\ar@{-}[dr]\\
      *=0{\bullet}&&*=0{\bullet}\ar@{-}[d]\\
       &&*=0{\bullet}
          }
      &
        \xymatrix @C = 5pt @R = 5pt{
        &&&\snode\ar@{-}[dl]\ar@{-}[dr]\\
       &&\snode\ar@{-}[dl]\ar@{-}[drr]&&\scarre\\
       &\snode\ar@{-}[dl]\ar@{-}[dr]&&&\snode\ar@{-}[dl]\ar@{-}[dr]\\
      \scarre&&\scarre &\snode\ar@{-}[dl]\ar@{-}[dr]&&\scarre\\
       &&\scarre&&\scarre}
      \\\\ 
      \xymatrix @C = 5pt @R = 5pt{
       &*=0{\bullet}\ar@{-}[dl]\ar@{-}[dr]\\
      *=0{\bullet}\ar@{-}[d]&&*=0{\bullet}\\
      *=0{\bullet}\ar@{-}[d]\\
      *=0{\bullet}
      }
       &
         \xymatrix @C = 5pt @R = 5pt{
      &&&&\snode\ar@{-}[dl]\ar@{-}[drr]\\
       &&&\snode\ar@{-}[dl]\ar@{-}[dr]&&&\snode\ar@{-}[dl]\ar@{-}[dr]\\
       &&\snode\ar@{-}[dl]\ar@{-}[dr]&&\scarre&\scarre&&\scarre\\
       &\snode\ar@{-}[dl]\ar@{-}[dr]&&\scarre\\
      \scarre&&\scarre
                }
      & 
        \xymatrix @C = 5pt @R = 5pt{
        &*=0{\bullet}\ar@{-}[dl]\ar@{-}[dr]\\
      *=0{\bullet}\ar@{-}[d]&&*=0{\bullet}\ar@{-}[d]\\
      *=0{\bullet}&&  *=0{\bullet}
                     }
      & \xymatrix @C = 5pt @R = 5pt{ &&&\snode\ar@{-}[dl]\ar@{-}[drr]\\
       &&\snode\ar@{-}[dl]\ar@{-}[dr]&&&\snode\ar@{-}[dl]\ar@{-}[dr]\\ &
                                                                         \snode\ar@{-}[dl]\ar@{-}[dr]&&\scarre&
                                                                                                                \snode\ar@{-}[dl]\ar@{-}[dr]&&\scarre\\ \scarre &&\scarre&\scarre&&\scarre
                                                                                                                                                                                    }       
      \\\\ 
      \xymatrix @C = 5pt @R = 5pt{
       &*=0{\bullet}\ar@{-}[dl]\ar@{-}[dr]\\
      *=0{\bullet}&&*=0{\bullet}\ar@{-}[d]\\
       &&*=0{\bullet}\ar@{-}[d]\\
       &&*=0{\bullet}
          }
      &
        \xymatrix @C = 5pt @R = 5pt{
        &&&\snode\ar@{-}[dl]\ar@{-}[drr]\\
       &&\snode\ar@{-}[dl]\ar@{-}[dr]&&&\snode\ar@{-}[dl]\ar@{-}[dr]\\
       &\scarre&&\scarre&\snode\ar@{-}[dl]\ar@{-}[dr]&&\scarre\\
       &&&\snode\ar@{-}[dl]\ar@{-}[dr]&&\scarre\\
       &&\scarre&&\scarre
                   }
        & 
          \xymatrix @C = 5pt @R = 5pt{
          &&*=0{\bullet}\ar@{-}[dl]\ar@{-}[dr]\\
       &*=0{\bullet}\ar@{-}[dl]\ar@{-}[dr]&&*=0{\bullet}\\
      *=0{\bullet}&&*=0{\bullet}
                     }
      &
        \xymatrix @C = 5pt @R = 5pt{
        &&&\snode\ar@{-}[dl]\ar@{-}[drrr]\\
       &&\snode\ar@{-}[dl]\ar@{-}[drr]&&&&\snode\ar@{-}[dl]\ar@{-}[dr]\\
       &\snode\ar@{-}[dl]\ar@{-}[dr]&&&\snode\ar@{-}[dl]\ar@{-}[dr]&\scarre&&\scarre\\
      \scarre&&\scarre&\scarre&&\scarre
                                 }
      \\\\ 
      \xymatrix @C = 5pt @R = 5pt{
       &&*=0{\bullet}\ar@{-}[dl]\ar@{-}[dr]\\
       &*=0{\bullet}&&*=0{\bullet}\ar@{-}[dl]\ar@{-}[dr]\\
       &&*=0{\bullet}&&*=0{\bullet}
                        }
        &
          \xymatrix @C = 5pt @R = 5pt{
          &&\snode\ar@{-}[dl]\ar@{-}[drr]\\
       &\snode\ar@{-}[dl]\ar@{-}[dr]&&&\snode\ar@{-}[dl]\ar@{-}[drr]\\
      \scarre&&\scarre&\snode\ar@{-}[dl]\ar@{-}[dr]&&&\snode\ar@{-}[dl]\ar@{-}[dr]\\
       &&\scarre   &&\scarre&\scarre&&\scarre
                                       }
    \end{array}
  \end{displaymath}
 
  \caption[Motzkin trees and their slanting trees]{$9$ Motzkin trees with
    $4$ edges and their slanting trees}
  \label{fig:truc}
\end{figure}

Following Dulucq and Penaud,
I represent \emph{Motzkin trees} as specific \emph{binary trees} in which
leaves~$\Box$ are added.  In such binary trees, only the three first
configurations below are allowed and the fourth and rightmost one is not.
\begin{displaymath}
  \xymatrix @C = 4pt @R = 4pt{
    &\snode{~}\ar@{-}[dl]\ar@{-}[dr]\\
    \snode&&\snode
  }
  \qquad
  \xymatrix @C = 4pt @R = 4pt{
    &\snode{~}\ar@{-}[dl]\ar@{-}[dr]\\
    \snode&&\scarre{~}
  }
  \qquad
  \xymatrix @C = 4pt @R = 4pt{
    &\snode{~}\ar@{-}[dl]\ar@{-}[dr]\\
    \scarre&&\scarre{~}
  }
  \qquad
  \xymatrix @C = 4pt @R = 4pt{
    \ar@{=}[drrrr]&&\snode{~}\ar@{-}[dl]\ar@{-}[dr]&&\ar@{=}[dllll]\\
    &\scarre{~}&&\snode&
  }
\end{displaymath}
The first configuration corresponds to a binary node, the second
configuration corresponds to a unary node and the third configuration
corresponds to an end node in the classical presentation (for instance in
Figure~\ref{fig:truc}).  I call such trees \emph{slanting trees} or
\emph{binary slanting trees}, from the french \emph{arbres binaires
  penchés}.

Figure~\ref{fig:truc} shows the $9$ Motzkin trees with $4$ edges and their
corresponding slanting trees.  Let us label each node of a slanting tree
with a number between $1$ and $2m+1$, where $m$ is the number of internal
nodes of the slanting tree. Let us call such a labeled tree a
\emph{labeled slanting tree}.  Now consider labeled slanting trees with
one marked leaf.  Let us call it a \emph{leaf-marked slanting tree}.
Below there is a labeled slanting tree of size $4$ and a leaf-marked
labeled slanting tree, where the mark is on the leaf labeled $8$.

\begin{displaymath}
  \xymatrix@C = 8pt @R = 8pt{
    &&&\noeud{3}\\
    && \node{5}\ar@{-}[dl]\ar@{-}[drr]&&\carre{4}\\
    &\noeud{7}&&&\noeud{1}\\
    \carre{6}&&\carre{8}&\carre{0}&&\carre{2}
  }
  \qquad \qquad
  \xymatrix@C = 8pt @R = 8pt{
    &&&\noeud{3}\\
    && \node{5}\ar@{-}[dl]\ar@{-}[drr]&&\carre{4}\\
    &\noeud{7}&&&\noeud{1}\\
    \carre{6}&&\CARRE{8}&\carre{0}&&\carre{2}
  }
\end{displaymath}
This corresponds to the vector
\begin{math}
  [3, 0, 2, 5 , 4, 7, 1, 6, 8].
\end{math}

How nodes and leaves are labeled by numbers will be explained below and is
essentially like binary trees. Like binary trees, just notice that
\emph{internal nodes} have odd labels and \emph{leaves} have even labels.
From now on, let us forget the labels, but let us mark one of the leaves.
In such trees with a marked leaf, we can distinguish $7$~general patterns
of subtrees containing the marked leaf (Figure~\ref{fig:pat} first
column). The marked leaf is denoted by a star in a square,
namely~$\xymatrix{\sCARRE}$.  In the first group of $4$ slanting trees,
there are the patterns where the marked leaf is a right child and in the
second group of $3$ slanting trees, there are the patterns where the
marked leaf is a left child, hence, due to the constraints on slanting
trees, the other child (a right child) is a leaf as well.

Let us call \emph{node-marked} a slanting tree in which one internal node
is marked.  Let us call \emph{marked} tree, a slanting tree in which
either a leaf or an internal node is marked.

\subsection*{How many leaves in a Motzkin tree?}
\label{how}

The slanting tree associated with a Motzkin tree of size $n$ (number of
its edges) has $n+2$ leaves. This can be shown by induction.
\begin{description}
\item[Basic case:] If the Motzkin tree is $\bullet$, its size is $0$ and
  its associated slanting tree
  \raisebox{6pt}{$\xymatrix@C=2pt@R=2pt{&\snode\ar@{-}[dl]\ar@{-}[dr]\\\scarre&&\scarre}$}
  has $2$ leaves.
  
\item[Adding a unary node:] Assume we add a unary node above a Motzkin
  tree $t$ of size $n$, this yields a Motzkin tree $t'$ of size $n+1$
  . The slanting tree associated with $t'$ has $n+2$ leaves (the number of
  the leave of the slanting tree $u$ associated with $t$) plus a new one
  added, then all together $n+3$.
  \begin{eqnarray*}
    t &\to& \raisebox{6pt}{$\xymatrix@C=2pt@R=4pt{\snode\ar@{-}[d]\\t}$} \\
    u &\to& \raisebox{6pt}{$\xymatrix@C=2pt@R=2pt{&\snode\ar@{-}[dl]\ar@{-}[dr]\\u &&\scarre}$}
  \end{eqnarray*}
\item[Adding a binary node:] Assume we add a binary node above two Motzkin
  trees $t_1$ and $t_2$ of size $n_1$ and $n_2$, this yields a Motzkin
  tree $t'$ of size $n_1+n_2+2$.  The slanting trees associated with
  \raisebox{6pt}{$\xymatrix@C=2pt@R=2pt{&\snode\ar@{-}[dl]\ar@{-}[dr]\\
      t_1 &&t_2}$} have $n_1+ n_2+4$ leaves.
\end{description}

\begin{figure}[!]
  \begin{center}
    \begin{tabular}[c]{l||c|c|c}
      &\textbf{leaf-marked} & \textbf{} & \textbf{Choice and}\\
      &\textbf{slanting trees} & \textbf{marked} & \textbf{node-marked}\\
      &&\textbf{slanting trees}&\textbf{slanting trees}\\
      \hline
      1.&
          \(\xymatrix@C=4pt @R=4pt{
                            &&&\ar@{.}[d]\\
      &&&\snode\ar@{-}[dl]\ar@{-}[dr]\\
      && \snode\ar@{-}[dl]\ar@{-}[dr]&&\sCARRE\\
      &\sTRIANGLE&&\sTRIANGLE
                    } \)%
                                        &
                                          \(\xymatrix@C=4pt @R=4pt{
      &&\ar@{.}[d]\\
      && \msnode\ar@{-}[dl]\ar@{-}[dr]\\
      &\sTRIANGLE&&\sTRIANGLE
                    } \)%

      & 
      \\
      2.&
          \(\xymatrix@C=4pt @R=4pt{
                            &&\ar@{.}[d]\\
      &&\snode\ar@{-}[dl]\ar@{-}[drr]\\
      & \snode\ar@{-}[dl]\ar@{-}[dr]&&&\snode\ar@{-}[dl]\ar@{-}[dr]\\
      \scarre&&\sCARRE&\sTRIANGLE&&\sTRIANGLE
                                    } \)
      &&
                     \baisse{15pt}{\textbf{LR , }} \(\xymatrix@C=4pt @R=4pt{
                                                 &\ar@{.}[d]\\
      &\msnode\ar@{-}[dl]\ar@{-}[dr]\\
      \sTRIANGLE&&\sTRIANGLE
                   }\)%
      \\
      3.&
          \(\xymatrix@C=4pt @R=4pt{
                            &&\ar@{.}[d]\\
      &&\snode\ar@{-}[dl]\ar@{-}[dr]\\
      & \snode\ar@{-}[dl]\ar@{-}[dr]&&\scarre\\
      \scarre&&\sCARRE
                } \)%
                            &
                              \(\xymatrix@C=4pt @R=4pt{
                                        &&\ar@{.}[d]\\
      && \snode\ar@{-}[dl]\ar@{-}[dr]\\
      &\sCARRE&&\scarre
                 } \)%
      &
      \\
      4.&
          \(\xymatrix@C=4pt @R=4pt{
                            &&&\ar@{.}[d]\\
      &&&\snode\ar@{-}[dl]\ar@{-}[drr]\\
      && \snode\ar@{-}[dl]\ar@{-}[dr]&&&\snode\ar@{-}[dl]\ar@{-}[dr]\\
      &\sTRIANGLE&&\sTRIANGLE&\scarre&&\sCARRE
                                        } \)
                               &\(\xymatrix@C=4pt @R=4pt{
                                                 &&&\ar@{.}[d]\\
      &&&\snode\ar@{-}[dl]\ar@{-}[drr]\\
      && \snode\ar@{-}[dl]\ar@{-}[dr]&&&\sCARRE\\
      &\sTRIANGLE&&\sTRIANGLE
                    } \)
      & \rouge{\baisse{15pt}{\textbf{RR , }} \(\xymatrix@C=4pt @R=4pt{
                            &\ar@{.}[d]\\
      &\msnode\ar@{-}[dl]\ar@{-}[dr]\\
      \sTRIANGLE&&\sTRIANGLE}\)}
      \\ &&&\\\hline&&&\\
      5.&
          \(\xymatrix@C=4pt @R=4pt{
                            &&\ar@{.}[d]\\
      &&\snode\ar@{-}[dl]\ar@{-}[drr]\\
      & \snode\ar@{-}[dl]\ar@{-}[dr]&&&\snode\ar@{-}[dl]\ar@{-}[dr]\\
      \sTRIANGLE&&\sTRIANGLE&\sCARRE&&\scarre
                                       } \)
      &&
                    \baisse{15pt}{\textbf{RL , }} \(\xymatrix@C=4pt @R=4pt{
                                                 &\ar@{.}[d]\\
      &\msnode\ar@{-}[dl]\ar@{-}[dr]\\
      \sTRIANGLE&&\sTRIANGLE}\)
                        
      \\ 
      6.&
          \(\xymatrix@C=4pt @R=4pt{
                            &&\ar@{.}[d]\\
      &&\snode\ar@{-}[dl]\ar@{-}[drr]\\
      & \snode\ar@{-}[dl]\ar@{-}[dr]&&&\snode\ar@{-}[dl]\ar@{-}[dr]\\
      \sCARRE&&\scarre&\sTRIANGLE&&\sTRIANGLE
                                    } \)
      &&
                        \baisse{15pt}{\textbf{LL , }} \(\xymatrix@C=4pt @R=4pt{
                                                 &\ar@{.}[d]\\
      &\msnode\ar@{-}[dl]\ar@{-}[dr]\\
      \sTRIANGLE&&\sTRIANGLE}\)
      \\
      7.&
          \(\xymatrix@C=4pt @R=4pt{
                            &&\ar@{.}[d]\\
      &&\snode\ar@{-}[dl]\ar@{-}[dr]\\
      & \snode\ar@{-}[dl]\ar@{-}[dr]&&\scarre\\
      \sCARRE&&\scarre
                } \)
                            & \raisebox{-15pt}{\dSmiley} ~
                              \(\xymatrix@C=4pt @R=4pt{
                                        &&\ar@{.}[d]\\
      && \snode\ar@{-}[dl]\ar@{-}[dr]\\
      &\scarre&&\sCARRE
                 } \)%
      &
      \\\hline
    \end{tabular}   
  \end{center}
  
  \caption{The $7$ patterns of leaf-marked slanting trees}
  \label{fig:pat}
\end{figure}

\subsection{A taxonomy of slanting trees}
\label{sec:taxo}

Recall Rémy's algorithm which consists in inserting, at a marked position
(internal node or leaf), a leaf in a marked tree (see
Figure~\ref{fig:Remy}).  After insertion, the formerly marked node becomes
unmarked and the inserted leaf becomes marked.  Here, since we are
interested in Motzkin trees, we insert a leaf on the right, above the
marked node in the marked slanting tree and like in \emph{Rémy's
  algorithm}, a leaf insertion on a marked slanting tree is performed, but unlike
\emph{Rémy's algorithm} the insertion is performed only on the right and a
leaf-marked slanting tree is produced.  This corresponds to what is done
to \emph{pattern1}, \emph{pattern3} and \emph{pattern4} in the middle
column of Figure~\ref{fig:pat}.  The leaf is inserted at a marked position
in the marked tree of the middle column producing the leaf-marked tree of
the left column.  But as we will see for the other patterns, there are
other ways to increase a slanting tree when it does not correspond to one
of these three patterns.

\section{The bijection}
\label{sec:bij}

Beside the fact one works on slanting trees with constraints, what makes
also the random generation of Motzkin trees trickier than Rémy's algorithm
is the structure of the holonomic equation:
\begin{displaymath}
  (n+2) M_n = (2n+1) M_{n-1} + 3 (n-1)M_{n-2}
\end{displaymath}
when compared to the equation:
\begin{displaymath}
  (n+1)  C_{n}= 2(2n-1) C_{n-1}
\end{displaymath}
First, if we use a construction based on that equation, a Motzkin tree of
size $n$ can be built from a Motzkin tree of size $n-1$ or from a Motzkin
tree of size $n-2$. Thus there are at least two cases to consider.
Actually $7$ cases as we will see, since the construction of a Motzkin
tree of size $n-1$ splits in $4$ cases and the construction of a Motzkin
tree of size $n-2$ splits in $3$ cases.  Notice that $M_n$ counts both the
number of Motzkin trees of size $n$ and the slanting trees with $n+2$
leaves.

\subsection*{Interpreting the holonomic equation}
\label{Interp}

We conclude that $(n+2)M_n$ counts the number of \emph{leaf-marked}
slanting trees of size~$n$, that $(2n+1)M_{n-1}$ counts the number of
\emph{ marked} slanting trees of size $n-1$ and that $(n-1)M_{n-2}$ counts
the number of \emph{node-marked} slanting trees of size $n-2$. Therefore
looking at the equation, we see that we should be able to build a
leaf-marked slanting tree of size $n$ from \underline{either} a marked
slanting tree of size $n-1$ \underline{or} from a pair made of an item
which can take one of three values and of a \emph{node-marked} slanting tree of
size $n-2$.  Let us see how Dulucq and Penaud propose to proceed.

\begin{figure}[!]
  \centering
  \begin{displaymath}
    \begin{array}[h!]{c|c|c}
      \hline
      \mathbf{1}&
                  \xymatrix@C=4pt @R=4pt{
      &&&\ar@{.}[d]\\
                &&&\snode\ar@{-}[dl]\ar@{-}[dr]\\
                && \snode\ar@{-}[dl]\ar@{-}[dr]&&\sCARRE\\
                &\sTRIANGLE&&\sTRIANGLE\\
                &~
                              } %
        &
          \xymatrix@C=4pt @R=4pt{
                &&\ar@{.}[d]\\
                && \msnode\ar@{-}[dl]\ar@{-}[dr]\\
                &\sTRIANGLE&&\sTRIANGLE
                              }%
      \\\hline
      \mathbf{3}&
                  \xymatrix@C=4pt @R=4pt{
      &&\ar@{.}[d]\\
                &&\snode\ar@{-}[dl]\ar@{-}[dr]\\
                & \snode\ar@{-}[dl]\ar@{-}[dr]&&\scarre\\
      \scarre&&\sCARRE
                } %
      &
        \xymatrix@C=4pt @R=4pt{
        &&\ar@{.}[d]\\
                && \snode\ar@{-}[dl]\ar@{-}[dr]\\
                &\sCARRE&&\scarre
                           }%
      \\\hline
      \mathbf{4}&
                  \xymatrix@C=4pt @R=4pt{
      &&&\ar@{.}[d]\\
                &&&\snode\ar@{-}[dl]\ar@{-}[drr]\\
                && \snode\ar@{-}[dl]\ar@{-}[dr]&&&\snode\ar@{-}[dl]\ar@{-}[dr]\\
                &\sTRIANGLE&&\sTRIANGLE&\scarre&&\sCARRE\\
                &~
                                                  } 
                &\xymatrix@C=4pt @R=4pt{
                &&&\ar@{.}[d]\\
                &&&\snode\ar@{-}[dl]\ar@{-}[drr]\\
                && \snode\ar@{-}[dl]\ar@{-}[dr]&&&\sCARRE\\
                &\sTRIANGLE&&\sTRIANGLE
                              } 
      \\\hline
      \mathbf{7}&
                  \xymatrix@C=4pt @R=4pt{
      &&\ar@{.}[d]\\
                &&\snode\ar@{-}[dl]\ar@{-}[dr]\\
                & \snode\ar@{-}[dl]\ar@{-}[dr]&&\scarre\\
      \sCARRE&&\scarre
                }
      & \raisebox{-15pt}{\dSmiley} ~
        \xymatrix@C=4pt @R=4pt{
        &&\ar@{.}[d]\\
                && \snode\ar@{-}[dl]\ar@{-}[dr]\\
                &\scarre&&\sCARRE
                           } %
      \\\hline
    \end{array}
  \end{displaymath}
  \caption{Contribution of marked slanting tree of size $n-1$.  $\dSmiley$
    for \emph{pattern7} marks a specific case explained in Section \emph{The
    lower part}}
  \label{fig:fstcontrib}
\end{figure}
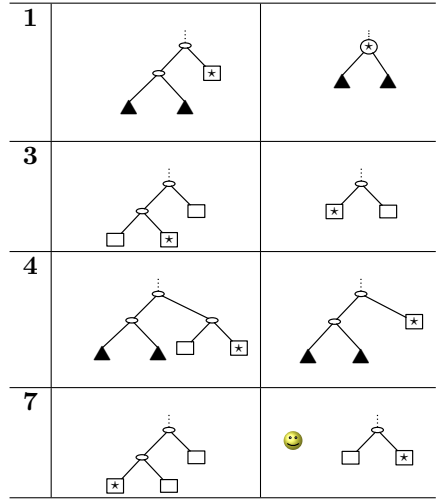
\subsection*{A taxonomy of leaf-marked slanting trees}
\label{sec:taxolmslt}
Leaf-marked slanting trees can be sorted according to the position of
their mark.  This is done in the first column of Figure~\ref{fig:pat}.
This column has two parts.

\subsubsection*{The upper part}
\label{sec:upper}

In the upper part, we have four patterns in which the marked leaf is a
right child. Let us call them \emph{pattern1}, \emph{pattern2},
\emph{pattern3} and \emph{pattern4} (Figure~\ref{fig:pat}).  Three of them
\emph{pattern1}, \emph{pattern3} and \emph{pattern4} are obtained by
Rémy's right insertion of a leaf in a marked slanting tree. The other
\emph{pattern2} is not. Indeed if the marked leaf is removed, the tree
that is obtained has a leaf on the left and a node on the right, which is
forbidden. This pattern will be obtained another way.

\subsubsection*{The lower part}
\label{sec:lower}

In the lower part, there are three patterns which correspond to the case
where the marked leaf is a left child, hence the sibling of a leaf
(Figure~\ref{fig:sndcontrib}); \emph{pattern7} cannot be obtained by a
right insertion of a leaf (Figure~\ref{fig:pat}).  I annotate it with a
$\dSmiley$.  But Dulucq and Penaud noticed that pattern
\raisebox{6pt}{$\xymatrix@C=2pt@R=2pt{&\snode\ar@{-}[dl]\ar@{-}[dr]\\\scarre
    &&\sCARRE}$} among marked slanting trees is not taken into
consideration. Thus they propose to associate this pattern
\raisebox{6pt}{$\xymatrix@C=2pt@R=2pt{&\snode\ar@{-}[dl]\ar@{-}[dr]\\\scarre
    &&\sCARRE}$} with \emph{pattern7}, as shown in Figure~\ref{fig:pat}.

\subsection*{The bijection by cases}
\label{byCases}

\subsubsection*{Case $(2n+1)M_{n-1}$}
\label{sec:Case1}
The previously explained contribution to leaf-marked slanting trees of
size $n$ from marked slanting tree of size $n-1$ is summarized in
Figure~\ref{fig:fstcontrib}.  All the patterns of marked slanting trees
are taken into account.

\subsubsection*{Case $3(n-1)M_{n-2}$}
\label{sec:case2}
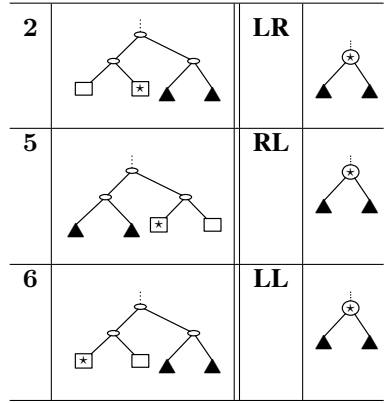
\begin{figure}[!]
  \centering
  \begin{displaymath}
    \begin{array}[h!]{c|c||c|c}
      \hline
      \mathbf{2}&
                  \raisebox{8pt}{$\xymatrix@C=4pt @R=4pt{
      &&\ar@{.}[d]\\
                &&\snode\ar@{-}[dl]\ar@{-}[drr]\\
                & \snode\ar@{-}[dl]\ar@{-}[dr]&&&\snode\ar@{-}[dl]\ar@{-}[dr]\\
      \scarre&&\sCARRE&\sTRIANGLE&&\sTRIANGLE\\~
                                         }$} 
                &\textbf{LR}
                                              & \xymatrix@C=4pt @R=4pt{
                                                &\ar@{.}[d]\\
                &\msnode\ar@{-}[dl]\ar@{-}[dr]\\
      \sTRIANGLE&&\sTRIANGLE
                        }%
      \\\hline
      \mathbf{5}&
                  \xymatrix@C=4pt @R=4pt{
      &&\ar@{.}[d]\\
                &&\snode\ar@{-}[dl]\ar@{-}[drr]\\
                & \snode\ar@{-}[dl]\ar@{-}[dr]&&&\snode\ar@{-}[dl]\ar@{-}[dr]\\
      \sTRIANGLE&&\sTRIANGLE&\sCARRE&&\scarre\\~
                                                 } 
                &
                  \textbf{RL}
                                              &
                                                \xymatrix@C=4pt @R=4pt{
                                                &\ar@{.}[d]\\
                &\msnode\ar@{-}[dl]\ar@{-}[dr]\\
      \sTRIANGLE&&\sTRIANGLE}
      \\\hline
      \mathbf{6}&
                  \xymatrix@C=4pt @R=4pt{
      &&\ar@{.}[d]\\
                &&\snode\ar@{-}[dl]\ar@{-}[drr]\\
                & \snode\ar@{-}[dl]\ar@{-}[dr]&&&\snode\ar@{-}[dl]\ar@{-}[dr]\\
      \sCARRE&&\scarre&\sTRIANGLE&&\sTRIANGLE\\~
                                         }
                &
                  \textbf{LL}
                                              &\xymatrix@C=4pt @R=4pt{
                                                &\ar@{.}[d]\\
                &\msnode\ar@{-}[dl]\ar@{-}[dr]\\
      \sTRIANGLE&&\sTRIANGLE}\\\hline
    \end{array}
\end{displaymath}
  \caption[Contribution of $n-2$]{Contribution of node-marked slanting tree of size $n-2$}
  \label{fig:sndcontrib}
\end{figure}

Let us now look at the three remaining patterns: \emph{pattern2},
\emph{pattern5} and \emph{pattern6}; forming the lines of
Figure~\ref{fig:sndcontrib}.  Those three patterns have the same model,
namely an internal node with two children: one child is an internal node and the
other child is an internal node whose children are two leaves, one of which
is marked, the other is not.  Depending on the position of the marked
leaf, we distinguish three cases.
\begin{itemize}
\item \textbf{LL} corresponds to the case where the marked node is on the left of the top node and on the left of its parent node.
\item \textbf{LR} corresponds to the case where the marked node is on the left of the top node and on the right of its parent node.
\item \textbf{RL} corresponds to the case where the marked node is on the right of the top node and on the left of its parent node.
\end{itemize}

One notices that there is no case \textbf{RR}, because this would
correspond to \emph{pattern4} considered in the previous section.  As a
matter of fact, the three cases \textbf{LL}, \textbf{LR} and \textbf{RL}
correspond to the multiplicative factor $3$ in the holonomic equation.

\subsubsection*{Forgetting the marks on leaves}
\label{sec:eras}
As Rémy noticed for binary trees, since we generate the leaf-marked
slanting trees of size $n$ uniformly, we also get a uniform distribution of
slanting trees of size $n$.  Thus we can forget the marks, which we do in
the concrete algorithm.
\begin{figure}[!]
  \par\noindent\rule{\textwidth}{0.4pt}
  \centering
\ifMint
\begin{minted}{haskell}
rMt :: Int -> Int -> Gen (Vector Int)
rMt seed 0 = do put (mkStdGen seed)
                return $ initialVector // [(0,1),(1,0),(2,2)]
rMt seed 1 = do put (mkStdGen seed)
                return $ initialVector // [(0,1),(1,3),(2,0),(3,2),(4,4)]
rMt seed n =  
   do generator <- get
      let (rand, newGenerator) = randomR (0::Double,1) generator
      put newGenerator
      case oracle n rand  of
        True -> case1 seed n
        False -> case2 seed n

case1 seed n =
   do generator <- get
      let (rand, newGenerator) = randomR (0::Double,1) generator
          k = floor (rand * (fromIntegral (2*n)))
      v <- rMt seed (n-1)
      put newGenerator
      case odd k || odd (v!k) || odd (v!(k-1)) of
       True -> return $ v // [(k,2*n+1),(2*n+1,v!k),(2*n+2,2*n+2)]
       False -> return $ v // [(k-1,2*n+1),(2*n+1,v!(k-1)),(2*n+2,2*n+2)]
       
case2 seed n =
    do generator <- get
       let (rand, newGenerator) = randomR (0::Double,1) generator
           r = floor (rand * (fromIntegral (3*n-6)))
           k = r `div` 3
           c = r `rem` 3
       v <- rMt seed (n-2)
       put newGenerator
       case c < 2 of
         True -> return $ v // [(2*k+1,2*n-1),(2*k+2,2*n+1),
                                (2*n-1,2*n),(2*n,2*n+2),
                                (2*n+1,v!(2*k+1)),(2*n+2,v!(2*k+2))]
         False -> return $ v // [(2*k+1,2*n-1),(2*k+2,2*n+1),
                                 (2*n-1,v!(2*k+1)),(2*n,v!(2*k+2)),
                                 (2*n+1,2*n),(2*n+2,2*n+2)]
\end{minted}
  \else
\begin{lstlisting}
rMt :: Int -> Int -> Gen (Vector Int)
rMt seed 0 = do put (mkStdGen seed)
                return $ initialVector // [(0,1),(1,0),(2,2)]
rMt seed 1 = do put (mkStdGen seed)
                return $ initialVector // [(0,1),(1,3),(2,0),(3,2),(4,4)]
rMt seed n =  
   do generator <- get
      let (rand, newGenerator) = randomR (0::Double,1) generator
      put newGenerator
      case oracle n rand  of
        True -> case1 seed n
        False -> case2 seed n

case1 seed n =
   do generator <- get
      let (rand, newGenerator) = randomR (0::Double,1) generator
          k = floor (rand * (fromIntegral (2*n)))
      v <- rMt seed (n-1)
      put newGenerator
      case odd k || odd (v!k) || odd (v!(k-1)) of
       True -> return $ v // [(k,2*n+1),(2*n+1,v!k),(2*n+2,2*n+2)]
       False -> return $ v // [(k-1,2*n+1),(2*n+1,v!(k-1)),(2*n+2,2*n+2)]
       
case2 seed n =
    do generator <- get
       let (rand, newGenerator) = randomR (0::Double,1) generator
           r = floor (rand * (fromIntegral (3*n-6)))
           k = r `div` 3
           c = r `rem` 3
       v <- rMt seed (n-2)
       put newGenerator
       case c < 2 of
         True -> return $ v // [(2*k+1,2*n-1),(2*k+2,2*n+1),
                                (2*n-1,2*n),(2*n,2*n+2),
                                (2*n+1,v!(2*k+1)),(2*n+2,v!(2*k+2))]
         False -> return $ v // [(2*k+1,2*n-1),(2*k+2,2*n+1),
                                 (2*n-1,v!(2*k+1)),(2*n,v!(2*k+2)),
                                 (2*n+1,2*n),(2*n+2,2*n+2)]  
\end{lstlisting} 
  \fi
 \caption{Haskell program for random generation of Motzkin trees}
 \par\noindent\rule{\textwidth}{0.4pt}
\label{fig:program}
\end{figure}
\section{A  concrete  algorithm for random generation of Motzkin trees}
\label{sec:cMT}

The \textsf{Haskell} program of Figure~\ref{fig:program} presents the
algorithm for random generation of Motzkin trees.  In what follows, I make
no distinction between the algorithm and the program and I consider the
\textsf{Haskell} program as an executable specification.  The main
function is called \textsf{rMt} and returns an object of type \textsf{Gen
  (Vector~Int)} like \textsf{rbt} in Figure~\ref{fig:programRemy}.

Assume that there is a function \textsf{motzkin} that returns the $n^{th}$
Motzkin number.  Like for Rémy's algorithm, one represents a labeled
slanting tree by a vector.  In this vector, odd labels are for internal  and even labels are for leaves.  Notice that the algorithm preserves
two properties:
\begin{enumerate}
\item The vector codes a slanting tree.
\item The vector of a Motzkin tree of size $n$ has a length $2n+3$. 
\end{enumerate}

In order to choose which case to consider, namely $(2n+1)M_{n-1}$
(\texttt{case1}) or $3(n-1)M_{n-2}$ (\texttt{case2}), the algorithm
\textsf{rMt} requires a random value between $0$ and $(n+2)M_{n}$ which we
call~$r$.  If $r$ is less than or equal to $(2n+1)M_{n-1}$, we are in
\texttt{case1}, else we are in \texttt{case2}.  Said otherwise, given a
random value $c$ between $0$ and $1$ if
$c \le \frac{(2n+1)M{n-1}}{(n+2)M_n}$ we choose \texttt{case1}, if not we
choose \texttt{case2}.

More abstractly, let us forget Motzkin numbers and consider an
\emph{oracle} which, given a random number~$r$ between $0$ and $1$ and
an $n$ chooses between \texttt{case1} and \texttt{case2}. If the oracle
runs in constant time and returns a boolean according to the distribution
given by the above inequality then the algorithm has a linear time
complexity and returns a random Motzkin tree distributed according to the
above inequality.

\begin{itemize}
\item \texttt{case1}: one draws at random a leaf or an internal node in a
  slanting tree of size $n-1$. This means choosing at random an index $k$
  in the vector $v$. We get \emph{pattern7} if three conditions are
  fulfilled.
  \begin{enumerate}
  \item \textbf{The marked item, should be a right child}. This means that \textbf{$k$ is even}, since the left child of a node of index $2p+1$ is $2p+1$ and the right child of this node is $2p+2$. 
  \item \textbf{The marked item is a leaf}.  This means that \textbf{$v[k]$ is even},
    since leaves have even labels.  Notice that Haskell uses the notation
    \texttt{v!k} for our mathematical notation $v[k]$.
  \item \textbf{The sibling item of the marked item is a leaf} (a left child by the way).   This means that \textbf{$v[k-1]$ is even}.
  \end{enumerate}
  In this case ($k$ is even, $v[k]$ is even and
  $v[k-1]$ is even) one inserts a node and a leaf as shown by
  Figure~\ref{fig:fstcontrib}, which corresponds in the code to:
  \begin{alltt}
    v // [(k-1,2*n+1),(2*n+1,v!(k-1)),(2*n+2,2*n+2)] 
  \end{alltt}\vspace*{-20pt}
  In \textsf{Haskell}, the operator \textsf{//} updates vectors at once,
  it is called a \textsf{bulk update}. \texttt{(2*n+1,v!(k-1))} means that
  the left child is a new node, at index $2*n+1$, which points to the
  former value of \texttt{v!(k-1)}. The right child is a new leaf.

  The other cases (\emph{pattern1}, \emph{pattern3}, \emph{pattern4})
  correspond to cases when one of $k$, $v[k]$ or $v[k-1]$ is
  odd. The update is then
  \begin{alltt}
    v // [(k,2*n+1),(2*n+1,v!k),(2*n+2,2*n+2)]
  \end{alltt}\vspace*{-20pt}
which corresponds to the first lines of Figure~\ref{fig:fstcontrib}.
\item \texttt{case2}: In this case we consider a random node-marked
  slanting trees of size $n-2$ and a random values among \textbf{LR},
  \textbf{RL}, \textbf{LL}.  For that we draw a number $r$ between $0$
  and $3n-6$, from which $r \div 3$ gives a random number between $0$ and
  $n-2$ (a random node) and $r~ \mathsf{mode}~3$ yields a random number
  among $0$, $1$ and $2$. We notice that \textbf{LR} and \textbf{LL}
  correspond to the same transformation, while \textbf{RL} corresponds to
  another transformation. In each case one adds four nodes, with labels $2n-1$, $2n$, $2n+1$ and $2n+2$.
Thus,
  \begin{alltt}
v // [(2*k+1,2*n-1),(2*k+2,2*n+1),(2*n-1,2*n),(2*n,2*n+2),
                                  (2*n+1,v!(2*k+1)),(2*n+2,v!(2*k+2))]
  \end{alltt}\vspace*{-20pt}
  is the transformation for \textbf{LR} and \textbf{LL}
  and
  \begin{alltt}
v // [(2*k+1,2*n-1),(2*k+2,2*n+1),(2*n-1,v!(2*k+1)),(2*n,v!(2*k+2)),
                                  (2*n+1,2*n),(2*n+2,2*n+2)]    
  \end{alltt}\vspace*{-20pt}
  corresponds to \textbf{RL}.  We let the reader check that the code of Figure~\ref{fig:program}
  corresponds to the pictures of Figure~\ref{fig:sndcontrib}.
\end{itemize}

\begin{figure}[!]
\centering \ifMint
\begin{minted}{haskell}
oracleMotzkin :: Int -> Double -> Bool
oracleMotzkin n rand =
  let r = floor (rand * fromIntegral (fromIntegral (n+2) * (motzkin n)))
  in r <= (fromIntegral (2*n+1)) * (motzkin (n-1))
\end{minted}
 \else
\begin{lstlisting}
oracleMotzkin :: Int -> Double -> Bool
oracleMotzkin n rand =
  let r = floor (rand * fromIntegral (fromIntegral (n+2) * (motzkin n)))
  in r <= (fromIntegral (2*n+1)) * (motzkin (n-1))
\end{lstlisting}
  \fi
  \caption{The plain oracle}
  \label{fig:oraMotz}
\end{figure}
\begin{figure}[!]
\centering \ifMint
\begin{minted}{haskell}
oracleRatioM :: Int -> Double -> Bool
oracleRatioM n rand = case n <= 100004 of
    True -> ratioM!n <= rand
    False -> 0.66666666667 <= rand
\end{minted}
 \else
 \begin{lstlisting}
   oracleRatioM :: Int -> Double -> Bool
oracleRatioM n rand = case n <= 100004 of
    True -> ratioM!n <= rand
    False -> 0.66666666667 <= rand
\end{lstlisting}
  \fi
  \caption{The approximation oracle}
  \label{fig:ora2}
\end{figure}

\section{Linearity and oracle}
\label{sec:linearity}

The linearity of the algorithm depends on an oracle which should decide an
inequality in constant time. In the implementation is the plain implementation of
Figure~\ref{fig:oraMotz}, the complexity of the algorithm \texttt{rMt} is
not asymptotically linear, due to computations on big numbers.

However the inequality $rand < \frac{(2n+1)M{n-1}}{(n+2)M_n}$ of the
oracle can be approximated by a precomputed table which I call
\texttt{ratioM} (of actual size $10004$ in my case), and which is used in
the oracle in Figure~\ref{fig:ora2}. For sizes larger that this bound, I
take just the value $0.66666666667$, which is a good approximation, since
the fraction goes down to the limit~$\frac{2}{3}$.  The actual
implementation runs up to size $9 \times 10^{6}$ and never requires (in
the benchmarks) more than $10$ digits of precision (see next section).
But despite the asymptotic linearity is not guaranteed, the algorithm is
linear in practice.

For a better efficiency, the program \texttt{rMt} can run without
recursive calls. In this case, the stack of the calls is first computed
and the construction of the vector is performed by popping the stack. I
wrote a \textsf{Python} program, which generates random Motzkin trees of
size $9$ millions in less that $50$ seconds on a laptop.  In all the
benchmarks, I checked that the difference between the drawn $rand$ and the
fraction is always larger than $10^{-10}$. Hence, digits that are not $6$
are never checked, which shows that the oracle called
\textsf{oracleRatioM} corresponds to the oracle called
\textsf{oracleMotzkin} in this size interval $[0\,..~9\times 10^6]$ and runs in
constant time.

\begin{figure}[!]
  \centering
  \begin{displaymath}
    \begin{array}{c@{\quad}c@{\quad}c@{\quad}c@{\quad}c@{\quad}c}
      \xymatrix @C = 5pt @R = 9pt{
       &\snode\ar@{-}[dl]\ar@{=}[dr]\\%
        \scarre &&\snode\ar@{-}[dl]\ar@{=}[dr]\\ %
        &\scarre &&\snode\ar@{-}[dl]\ar@{-}[dr]\\ %
       &&\scarre &&\scarre}
 & %
 \xymatrix @C = 5pt @R = 9pt{
     &\snode\ar@{-}[dl]\ar@{=}[dr]\\ 
      \scarre &&\snode\ar@{-}[dl]\ar@{-}[dr]\\%
       &\scarre &&\snode\ar@{-}[dl]\ar@{-}[dr]\\ %
       &&\scarre &&\scarre} %
 & %
 \xymatrix @C = 5pt @R = 9pt{
                 &\snode\ar@{-}[dl]\ar@{-}[dr]\\ %
      \scarre &&\snode\ar@{-}[dl]\ar@{-}[dr]\\%
       &\scarre &&\snode\ar@{-}[dl]\ar@{-}[dr]\\%
       &&\scarre &&\scarre}
 & %
    \xymatrix @C = 5pt @R = 9pt{
                 &\snode\ar@{-}[dl]\ar@{-}[dr]\\ %
      \scarre &&\snode\ar@{-}[dl]\ar@{=}[dr]\\%
       &\scarre &&\snode\ar@{-}[dl]\ar@{-}[dr]\\%
       &&\scarre &&\scarre}       
 \\\\ %
   \xymatrix @C = 5pt @R = 9pt{
                 &\snode\ar@{-}[dl]\ar@{-}[dr]\\ %
      \scarre &&\snode\ar@{-}[dl]\ar@{-}[dr]\\%
       &\snode\ar@{-}[dl]\ar@{-}[dr]&&\scarre\\%
       \scarre &&\scarre}
  & %
    \xymatrix @C = 5pt @R = 9pt{
                 &\snode\ar@{-}[dl]\ar@{=}[dr]\\ %
      \scarre &&\snode\ar@{-}[dl]\ar@{-}[dr]\\%
       &\snode\ar@{-}[dl]\ar@{-}[dr]&&\scarre\\%
       \scarre &&\scarre}
&
 \xymatrix  @C = 5pt @R = 9pt{
       && \snode\ar@{-}[dl]\ar@{-}[dr] \\%
       &\snode\ar@{-}[dl]\ar@{=}[dr] && \scarre \\%
      \scarre &&\snode\ar@{-}[dl]\ar@{-}[dr]\\%
       &\scarre&&\scarre }
 & %
  \xymatrix  @C = 5pt @R = 9pt{
       && \snode\ar@{-}[dl]\ar@{-}[dr] \\%
       &\snode\ar@{-}[dl]\ar@{-}[dr] && \scarre \\%
      \scarre &&\snode\ar@{-}[dl]\ar@{-}[dr]\\%
       &\scarre&&\scarre
                  }
  \\\\ %
 \xymatrix  @C = 5pt @R = 9pt{
       &&& \snode\ar@{-}[dl]\ar@{-}[dr] \\%
       &&\snode\ar@{-}[dl]\ar@{-}[dr] && \scarre \\%
       &\snode\ar@{-}[dl]\ar@{-}[dr] && \scarre \\%
       \scarre&&\scarre
                 } %
                & %
 \xymatrix  @C = 5pt @R = 9pt{
       && \snode\ar@{-}[dl]\ar@{-}[drr] \\%
       & \snode\ar@{-}[dl]\ar@{-}[dr] &&& \snode\ar@{-}[dl]\ar@{-}[dr] \\%
        \scarre&&\scarre & \scarre&&\scarre
 } 
                 & %
 \xymatrix  @C = 5pt @R = 9pt{
       && \snode\ar@{-}[dl]\ar@{=}[drr] \\%
       & \snode\ar@{-}[dl]\ar@{-}[dr] &&& \snode\ar@{-}[dl]\ar@{-}[dr] \\%
        \scarre&&\scarre & \scarre&&\scarre
                                     }
    \end{array}
\end{displaymath}
\caption{The $11$ Schröder trees with $4$ leaves.}
\label{fig:Schr}
\end{figure}
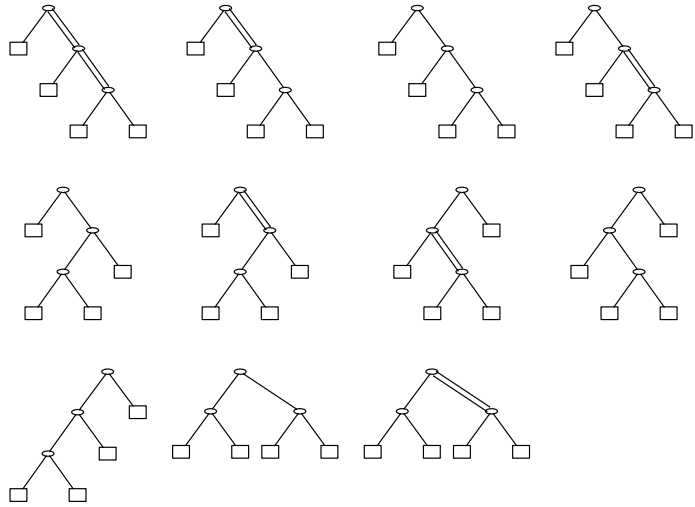

\section{Schröder trees}
\label{sec:schr}

In this section, I define Schröder trees and a proof by bijection of the
holonomic equation for defining numbers that count Schröder trees, aka
\emph{Schröder numbers} or \emph{Schröder-Hipparchus numbers}.  This proof
is the translation for Knuth's definition of Schröder trees of this due to
Foata and Zeilberger~\cite{DBLP:journals/jct/FoataZ97}.  From this proof I
derive a quasi linear algorithm for random generation of Schröder trees.

\subsection{Definition of Schröder trees}
\label{sec:def}

We take the definition of Knuth~\cite{KnuthVol4_4} § 7.2.1.6 (pp.~41):
\emph{''A \emph{Schröder tree} is a binary tree in which every nonnull
  right link is colored either white or black''}. We represent
\emph{black links} with~\raisebox{3pt}{$\xymatrix @C = 2.5pt @R = 5pt{\snode\ar@{-}[dr]\\&}$} and
\emph{white links} with \raisebox{3pt}{$\xymatrix @C = 1.5pt @R = 5pt{\snode\ar@{=}[dr]\\&}$}. But
unlike Knuth, we follow Foata and Zeilberger~\cite{DBLP:journals/jct/FoataZ97} and we say that a Schröder
tree has size $n$ if it has $n$ leaves, hence $n-1$ nodes.  For instance,
Schröder trees with $4$ leaves are given in Figure~\ref{fig:Schr}.

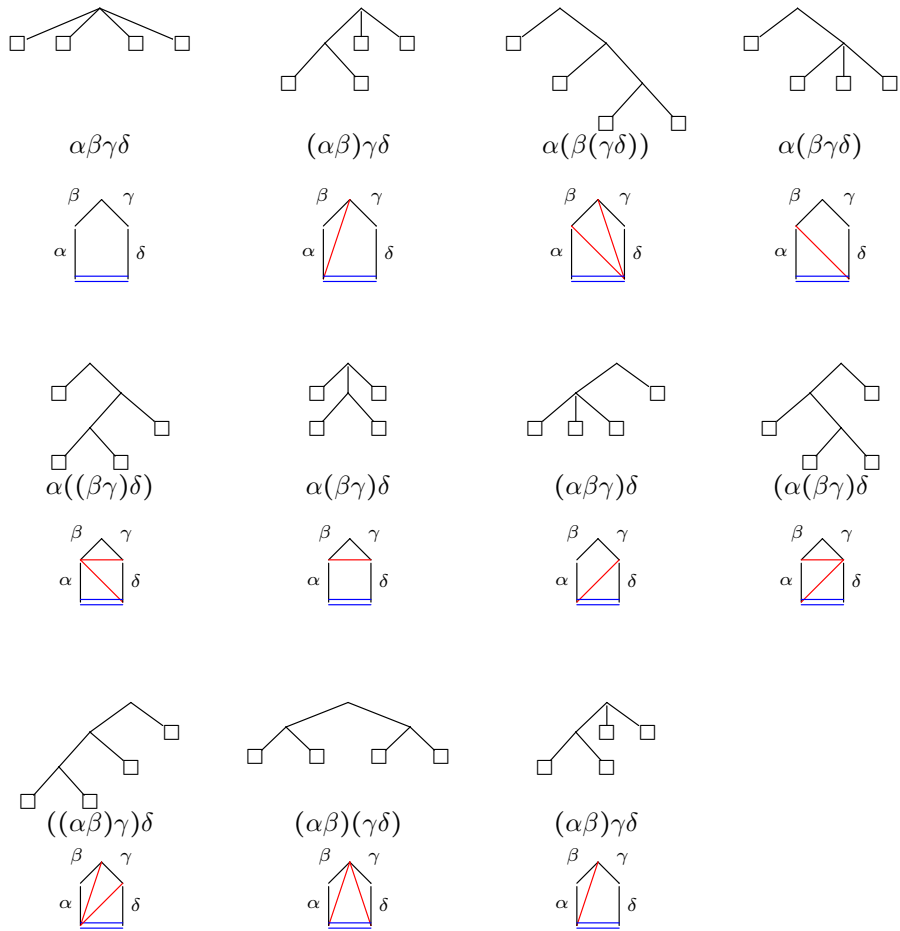
\begin{figure}[!]
  \centering
  \begin{displaymath}
  \begin{array}{c@{\qquad}c@{\qquad}c@{\qquad}c}
     \xymatrix  @M=0pt @C=10pt@R=10pt{
    &&\ar@{-}[dll]\ar@{-}[dl]\ar@{-}[dr]\ar@{-}[drr]\\
       \Box&\Box&&\Box&\Box }
    & %
      \xymatrix  @M=0pt @C=10pt@R=10pt{
    &&\ar@{-}[dl]\ar@{-}[d]\ar@{-}[dr]\\
    &\ar@{-}[dl]\ar@{-}[dr] &\Box&\Box\\
    \Box&&\Box }
    & %
      \xymatrix  @M=0pt @C=10pt@R=10pt{
        &\ar@{-}[dl]\ar@{-}[dr]\\
    \Box&&\ar@{-}[dl]\ar@{-}[dr]\\
    &\Box&&\ar@{-}[dl]\ar@{-}[dr]\\
    &&\Box&&\Box                             
             }
                           & %
    \xymatrix  @M=0pt @C=10pt@R=10pt{
     &\ar@{-}[dl]\ar@{-}[dr]\\
    \Box&&\ar@{-}[dl]\ar@{-}[d]\ar@{-}[dr]\\
    &\Box&\Box&\Box
                }
\\[40pt] 
    `a`b`g`d & (`a `b)`g `d & `a(`b(`g`d))& `a(`b`g`d)\\
\\[-5pt] 
 \xymatrix @M=0pt @C=10pt@R=10pt{
    &\ar@{-}[dl]_{\beta}  \ar@{-}[dr]^{\gamma} \\
    \ar@{-}[dd]_{\alpha}  &&\ar@{-}[dd]^{\delta}  \\\\
    \ar@{=}@[blue][rr]&&
  }
  &
  \xymatrix @M=0pt @C=10pt@R=10pt{
    &\ar@{-}[dl]_{\beta}  \ar@{-}[dr]^{\gamma} \ar@{-}@[red][dddl]\\
    \ar@{-}[dd]_{\alpha}  &&\ar@{-}[dd]^{\delta}  \\\\
    \ar@{=}@[blue][rr]&&
  }
  &
    \xymatrix @M=0pt @C=10pt@R=10pt{
    &\ar@{-}[dl]_{\beta}  \ar@{-}[dr]^{\gamma} \ar@{-} \ar@{-}@[red][dddr]\\
    \ar@{-}[dd]_{\alpha}  \ar@{-}@[red][ddrr] &&\ar@{-}[dd]^{\delta}  \\\\
    \ar@{=}@[blue][rr]&&
                              }
                              &       \xymatrix @M=0pt @C=10pt@R=10pt{
    &\ar@{-}[dl]_{\beta}  \ar@{-}[dr]^{\gamma} \ar@{-} \\
    \ar@{-}[dd]_{\alpha}  \ar@{-}@[red][ddrr] &&\ar@{-}[dd]^{\delta}  \\\\
    \ar@{=}@[blue][rr]&&
                              }
     \\[50pt] 
           \xymatrix  @M=0pt @C=8pt@R=8pt{
    &&\ar@{-}[dl]\ar@{-}[dr]\\
    &\Box&&\ar@{-}[dl]\ar@{-}[dr] \\
    &&\ar@{-}[dl]\ar@{-}[dr] &&\Box\\
    &\Box&&\Box}
   & %
   \xymatrix  @M=0pt @C=8pt@R=8pt{
    &\ar@{-}[dl]\ar@{-}[d]\ar@{-}[dr]\\
    \Box&\ar@{-}[dl]\ar@{-}[dr]&\Box\\
    \Box&&\Box}
    & %
    \xymatrix  @M=0pt @C=8pt@R=8pt{
     &&\ar@{-}[dl]\ar@{-}[dr]\\
    &\ar@{-}[dl]\ar@{-}[d]\ar@{-}[dr]&&\Box\\
    \Box&\Box&\Box
                }
 &         \xymatrix  @M=0pt @C=8pt@R=8pt{
        &&\ar@{-}[dl]\ar@{-}[dr]\\
    &\ar@{-}[dl]\ar@{-}[dr]&&\Box\\
    \Box&&\ar@{-}[dl]\ar@{-}[dr]\\
    &\Box&&\Box }
     \\[35pt]
    `a((`b  `g) `d)& `a(`b`g)`d & (`a`b`g)`d & (`a (`b`g)`d 

  \\[5pt] 

                 \xymatrix @M=0pt @C=8pt@R=8pt{
    &\ar@{-}[dl]_{\beta}  \ar@{-}[dr]^{\gamma} \\
    \ar@{-}[dd]_{\alpha}\ar@{-}@[red][rr]\ar@{-}@[red][rrdd]  &&\ar@{-}[dd]^{\delta}  \\\\
    \ar@{=}@[blue][rr]&&
  }
  &
  \xymatrix @M=0pt @C=8pt@R=8pt{
    &\ar@{-}[dl]_{\beta}  \ar@{-}[dr]^{\gamma} \\
    \ar@{-}[dd]_{\alpha} \ar@{-}@[red][rr] &&\ar@{-}[dd]^{\delta}  \\\\
    \ar@{=}@[blue][rr]&&
  }
  &
    \xymatrix @M=0pt @C=8pt@R=8pt{
    &\ar@{-}[dl]_{\beta}  \ar@{-}[dr]^{\gamma} \\
    \ar@{-}[dd]_{\alpha}   &&\ar@{-}[dd]^{\delta}\ar@{-}@[red][ddll] \\\\
    \ar@{=}@[blue][rr]&&
  }
  &
     \xymatrix @M=0pt @C=8pt@R=8pt{
    &\ar@{-}[dl]_{\beta}  \ar@{-}[dr]^{\gamma} \ar@{-} \\
    \ar@{-}[dd]_{\alpha}  \ar@{-}@[red][rr] &&\ar@{-}[dd]^{\delta} \ar@{-} \ar@{-}@[red][ddll] \\\\
    \ar@{=}@[blue][rr]&&}
\\[50pt] 
    \xymatrix  @M=0pt @C=8pt@R=8pt{
    &&&\ar@{-}[dl]\ar@{-}[dr]\\
    &&\ar@{-}[dl]\ar@{-}[dr]&&\Box \\
    &\ar@{-}[dl]\ar@{-}[dr] &&\Box\\
    \Box&&\Box}
   & %
   \xymatrix  @M=0pt @C=8pt@R=8pt{
    &&&\ar@{-}[dll]\ar@{-}[drr]\\
    &\ar@{-}[dl]\ar@{-}[dr]&&&&\ar@{-}[dl]\ar@{-}[dr] \\
    \Box&&\Box&&\Box&&\Box}
    & %
    \xymatrix  @M=0pt @C=8pt@R=8pt{
     &&\ar@{-}[dl]\ar@{-}[d]\ar@{-}[dr]\\
    &\ar@{-}[dl]\ar@{-}[dr]&\Box&\Box\\
    \Box&&\Box
               }
     \\[30pt] 
    ((`a`b)`g)`d & (`a`b) (`g`d)& (`a`b)`g`d
             
   \\ 

 \xymatrix @M=0pt @C=8pt@R=8pt{
    &\ar@{-}[dl]_{\beta}  \ar@{-}[dr]^{\gamma} \ar@{-}@[red][lddd]\\
    \ar@{-}[dd]_{\alpha} &&\ar@{-}[dd]^{\delta}  \\\\
    \ar@{=}@[blue][rr]\ar@{-}@[red][uurr]&&
  }
  &
  \xymatrix @M=0pt @C=8pt@R=8pt{
    &\ar@{-}[dl]_{\beta}  \ar@{-}[dr]^{\gamma} \ar@{-}@[red][dddl] \ar@{-}@[red][dddr] \\
    \ar@{-}[dd]_{\alpha} &&\ar@{-}[dd]^{\delta}  \\\\
    \ar@{=}@[blue][rr]&&
  }
  &
    \xymatrix @M=0pt @C=8pt@R=8pt{
    &\ar@{-}[dl]_{\beta}  \ar@{-}[dr]^{\gamma} \ar@{-}@[red][dddl]\\
    \ar@{-}[dd]_{\alpha}   &&\ar@{-}[dd]^{\delta} \\\\
    \ar@{=}@[blue][rr]&&
  }

          \end{array}
\end{displaymath}
\caption[$11$ trees with $4$ leaves and partitions of pentagons]{$11$
  Trees having no node of degree one and $4$ leaves, associated with\\
  $11$ partitions of pentagons}
\par\noindent\rule{\textwidth}{0.4pt}
  \label{fig:pentagons}
\end{figure}
Schröder trees are counted by numbers $S_n$, which form the sequence \href{https://oeis.org/A001003}{A001003} in
the \emph{online encyclopedia of integer sequences}~\cite{OEIS} and which fulfill the holonomic equations:
\begin{displaymath}
   (n+1)S_{n+1} - 3 (2n-1)S_n + (n-2) S_{n-1} = 0
 \end{displaymath}
 that we will use on the form 
\begin{displaymath}
  (n+1)S_{n+1} = 3 (2n-1)S_n - (n-2) S_{n-1}
\end{displaymath}
Notice that Foata and Zeilberger~\cite{DBLP:journals/jct/FoataZ97} use for the constructive proof:
\begin{equation}\label{eq:FZ}
   3 (2n-1)S_n =  (n+1)S_{n+1}  + (n-2) S_{n-1}
 \end{equation}
 Schröder trees count also \emph{ordinary trees with $n$ leaves and no
   node of degree one}. One ``shrinks'' the white edges, merging the nodes
 that they connects (see~\cite{KnuthVol4_4} Exercise~66). For trees with
 $4$ leaves, the reader can see the bijection in the lines of
 Figure~\ref{fig:Schr} and in the odd lines of Figure~\ref{fig:pentagons}.
 Trees in correspondence are on the same row.  Schröder numbers count also
 \emph{partitions of polygons}~\cite{etherington40}. The bijection between
 ordinary trees with $4$ leaves and no node of degree one and partitions
 of pentagons (polygons with $5$ edges) is illustrated by
 Figure~\ref{fig:pentagons}.  First one associates with each tree a
 parenthesized expression in $`a$, $`b$, $`g$ and $`d$, in this order.
 Then each parenthesized expression is associated with a unique partition
 of the pentagon. For this purpose, a basic edge
 $\!\!\xymatrix@C=12pt{\ar@{=}@[blue][r]&}\!\!$ is chosen and other edges
 are named clockwise by letters $`a$, $`b$, $`g$, $`d$.  If one proceeds
 starting at edge $`a$, each diagonal closes a polygon (with less edges)
 in which the other edges have been already named by an expression.
 Therefore one can associate a partition with each expression as this is
 presented in Figure~\ref{fig:pentagons}.  One can process the same way
 for hexagons, heptagons etc.
 
Didier Arquès and Alain Giorgetti have shown that \emph{planar rooted hypergraphs with
   vertices only of the outer face} are also counted by Schröder
 numbers~\cite{DBLP:journals/dm/ArquesG00}.

\subsection{Foata and Zeilberger bijection}
In identity~(\ref{eq:FZ}):
\begin{itemize}
\item $(2n-1) S_n$ counts marked Schröder trees of size $n$, where nodes
  and leaves can be marked, therefore $3 (2n-1) S_n$, counts pairs made of
  one of three values and of a marked Schröder tree,
\item $(n+1)S_{n+1}$ counts leaf-marked Schröder trees of size $n+1$,
\item $(n-2)S_{n-1}$ counts node-marked Schröder trees of size $n-1$.
\end{itemize}
In order to prove identity~(\ref{eq:FZ}), Foata and Zeilberger build a
one-to-one function which associates, with a pair of a number $0$, $1$ or
$2$ (coded by Foata and Zeilberger as $L_1$, $L_2$ and $R_1$) and a marked
Schröder tree of size $n$, either a leaf-marked Schröder tree of size
$n+1$ or a node-marked Schröder tree of size $n-1$.

\begin{figure}[!]
  \centering
  \begin{displaymath}
      \xymatrix@C=3pt @R=3pt{
      &&&\ar@{.}[d]\\
      &&&\snode\ar@{-}[dl]\ar@{-}[dr]\\
      &&\sCARRE&&*=<3pt>{\circ}\ar@{-}[ddll]\ar@{-}[ddrr]\\
      &&&&\mathsf{T}\\
      &&*=<1pt>{}\ar@{-}[rrrr]&&&&*=<1pt>{} }
    \qquad
    \xymatrix@C=3pt @R=3pt{
      &&&\ar@{.}[d]\\
      &&&\snode\ar@{-}[dl]\ar@{=}[dr]\\
      &&\sCARRE&&*=<3pt>{\circ}\ar@{-}[ddll]\ar@{-}[ddrr]\\
      &&&&\mathsf{T}\\
      &&*=<1pt>{}\ar@{-}[rrrr]&&&&*=<1pt>{} }
    \qquad
   \xymatrix@C=3pt @R=3pt{
      &&&\ar@{.}[d]\\
      &&&\snode\ar@{-}[dl]\ar@{-}[dr]\\
      &&*=<3pt>{\circ}\ar@{-}[ddll]\ar@{-}[ddrr]&&\sCARRE\\
      &&\mathsf{T}\\
      *=<1pt>{}\ar@{-}[rrrr]&&&&*=<1pt>{} }
  \end{displaymath}
  \caption{Insertion of a leaf in Foata and Zeilberger bijection}
  \label{fig:FZb}
\end{figure}
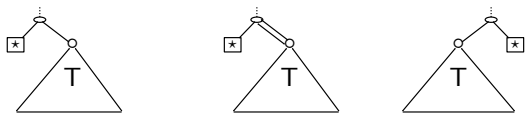

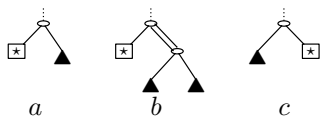
\begin{figure}[!]
  \centering
     \begin{displaymath}
       \begin{array}{c@{\qquad}c@{\qquad}c}
     \xymatrix@C=4pt @R=4pt{
      &\ar@{.}[d]\\
      &\snode \ar@{-}[dl]\ar@{-}[dr]\\
      \sCARRE&&\sTRIANGLE
    }
    &
     \xymatrix@C=4pt @R=4pt{
      &\ar@{.}[d]\\
      &\snode \ar@{-}[dl]\ar@{=}[dr]\\
      \sCARRE&&\snode\ar@{-}[dl]\ar@{-}[dr]\\
      &\sTRIANGLE&&\sTRIANGLE
    }
    &
     \xymatrix@C=4pt @R=4pt{
      &\ar@{.}[d]\\
      &\snode \ar@{-}[dl]\ar@{-}[dr]\\
      \sTRIANGLE&&\sCARRE
                   }
         \\[30pt]
         a & b & c
       \end{array}
  \end{displaymath}
  \caption[3 impossible]{Three impossible left insertions of leaves}
  \label{fig:ThreeImp}
\end{figure}

Foata and Zeilberger bijection~\cite{DBLP:journals/jct/FoataZ97} that proves identity~(\ref{eq:FZ}) is based
on constructions similar to Rémy's insertion (Figure~\ref{fig:Remy}), with
the difference that there are two left insertions, one that corresponds to
a black link and one that corresponds to a white link
(Figure~\ref{fig:FZb}).
Due to the constraints on white links in Schröder trees, there are three
patterns (Figure~\ref{fig:ThreeImp}), with impossible insertion, when one
tries to insert a left leaf to become a left sibling of a leaf connected
with a white link. Two patterns ($a$ and $c$
in Figure~\ref{fig:ThreeImp}) can be ''recovered''. The third pattern ($b$
in in Figure~\ref{fig:ThreeImp}) is shrunk toward a Schröder tree of size
$n-1$. This will correspond to a \textsf{fail} in the algorithm.

Similarly there are two unreachable cases by left or right insertion.
They are patterns which are the results of a right or left insertion on a
leaf which are a right child by a white link. 
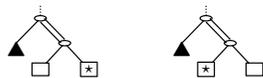
\begin{figure}[!]
  \begin{displaymath}
  \xymatrix@C=3pt@R=3pt{
     &&\ar@{.}[d]\\
      &&\snode\ar@{-}[dl]\ar@{=}[dr]\\
      &\sTRIANGLE&&\snode\ar@{-}[dr]\ar@{-}[dl]\\
      &&\scarre&&\sCARRE
    }
    \qquad
      \xymatrix@C=3pt@R=3pt{
     &&\ar@{.}[d]\\
      &&\snode\ar@{-}[dl]\ar@{=}[dr]\\
      &\sTRIANGLE&&\snode\ar@{-}[dr]\ar@{-}[dl]\\
      &&\sCARRE&&\scarre
    }
  \end{displaymath}
  \caption{Two unreachable}
  \label{fig:TwoUnreach}
\end{figure}

Let us call $L_1$, $L_2$ and $R_1$ the three labels. 
Therefore Figure~\ref{fig:FandZ} gives the correspondence between pairs of a label from $\{L_1, L_2, R_2\}$ and of a marked Schröder tree, with either a leaf-marked tree of size $n+1$ or a node marked tree of size $n-1$.

\begin{description}
\item[Cases $L_1$ and $R_1$.] One inserts a leaf with a black link. This is exactly like Rémy's insertion.
\item[Case $L_2$.] This case deals with a left insertion with a white right link.
  \begin{description}
  \item[First line:] The mark is on a node. The left insertion with a white right link is possible. 
  \item[Second line:] The mark is on a left leaf and the right link is
    black. The left insertion yields a forbidden pattern (a right white
    link toward a leaf). But by twisting the tree and swapping the colors
    of the right links, one reaches the first unreachable pattern of
    Figure~\ref{fig:TwoUnreach}.
    
  \item[Third line:] The mark is on a left leaf and the right link is
    white.  Then the left child is not a leaf. The left insertion is
    forbidden as well. Therefore by removing the left leaf one gets a node
    marked tree of size $n-1$.
  \item[Fourth line:] The mark is on a right leaf.  The left insertion
    yields a forbidden pattern, but by swapping the colors of the right link one gets
    the second unreachable pattern of Figure~\ref{fig:TwoUnreach}.
  \end{description}
\end{description}

\begin{figure}[!]
  \centering
  \begin{displaymath}
    \begin{array}{l|c@{\qquad\qquad}c@{\qquad\qquad}c}
      \raisebox{8pt}{$\mathbf{L_1}$} & \raisebox{35pt}{\xymatrix@C=3pt@R=3pt{&\ar@{.}[d] \\ &*=<3pt>{\scriptstyle \ostar}\ar@{.}[dl]\ar@{.}[dr]\\
       *=<1pt>{}\ar@{.}[rr]&&*=<1pt>{} }}&& %
      \stackrel{\xymatrix@C=5pt@R=6pt{&\ar@{.}[d]\\&\snode\ar@{-}[dl]\ar@{-}[dr]\\\sCARRE&&*=<3pt>{\scriptstyle \ostar}\ar@{.}[dl]\ar@{.}[dr]\\ &*=<1pt>{}\ar@{.}[rr]&&*=<1pt>{}}}{~} %
      \\\hline
      \raisebox{8pt}{$\mathbf{L_2}$} & \stackrel{\xymatrix@C=5pt@R=5pt{&\ar@{.}[d]\\ &\msnode\ar@{-}[dl]\ar@{-}[dr]\\\sTRIANGLE&&\sTRIANGLE}}{~} &&%
      \raisebox{25pt}{\xymatrix@C=5pt@R=5pt{&\ar@{.}[d]\\&\snode\ar@{-}[dl]\ar@{=}[dr]\\\sCARRE&&\snode\ar@{-}[dl]\ar@{-}[dr]\\ &\sTRIANGLE&&\sTRIANGLE}} %
      \\ 
      & \stackrel{\xymatrix@C=5pt@R=5pt{&\ar@{.}[d]\\ &\snode\ar@{-}[dl]\ar@{-}[dr]\\\sCARRE&&\sTRIANGLE}}{~} %
      & \stackrel{\xymatrix@C=5pt@R=5pt{&&\ar@{.}[d]\\ &&\snode\ar@{-}[dl]\ar@{-}[dr]\\ &\snode\ar@{-}[dl]\ar@{=}[dr]&&\sTRIANGLE\\ \sCARRE&&\scarre}}{~}
      \hspace*{70pt} \raisebox{20pt}{$\rla\quad \dua$}
      & \stackrel{\xymatrix@C=5pt@R=5pt{&\ar@{.}[d]\\ &\snode\ar@{-}[dl]\ar@{=}[dr]\\\sTRIANGLE &&\snode\ar@{-}[dl]\ar@{-}[dr]\\ &\scarre&&\sCARRE}}{~}
      \\ 
      & \stackrel{\xymatrix@C=5pt@R=5pt{&\ar@{.}[d]\\ &\snode\ar@{-}[dl]\ar@{=}[dr]\\\sCARRE&&\snode\ar@{-}[dl]\ar@{-}[dr]\\ &\sTRIANGLE&&\sTRIANGLE}}{~}
      & \hspace*{80pt}\raisebox{20pt}{$\searrow$}
      & \stackrel{\xymatrix@C=5pt@R=5pt{&\ar@{.}[d]\\ &\msnode\ar@{-}[dl]\ar@{-}[dr]\\\sTRIANGLE&&\sTRIANGLE}}{~}
      \\[5pt] 
      &\stackrel{\xymatrix@C=5pt@R=5pt{&\ar@{.}[d]\\ &\snode\ar@{-}[dl]\ar@{-}[dr]\\\sTRIANGLE&&\sCARRE}}{~}
      & \stackrel{\xymatrix@C=5pt@R=5pt{&\ar@{.}[d]\\ &\snode\ar@{-}[dl]\ar@{-}[dr]\\\sTRIANGLE&&\snode\ar@{-}[dl]\ar@{=}[dr]\\ &\sCARRE&&\scarre}}{~} 
      \qquad\qquad  \raisebox{20pt}{$\dua$}
      & \stackrel{\xymatrix@C=5pt@R=5pt{&\ar@{.}[d]\\ &\snode\ar@{-}[dl]\ar@{=}[dr]\\\sTRIANGLE&&\snode\ar@{-}[dl]\ar@{-}[dr]\\ &\sCARRE&&\scarre}}{~}          
      \\\hline
      \raisebox{8pt}{$\mathbf{R_1}$} &
     \raisebox{35pt}{\xymatrix@C=3pt@R=3pt{&\ar@{.}[d] \\ &*=<3pt>{\scriptstyle \ostar}\ar@{.}[dl]\ar@{.}[dr]\\ *=<1pt>{}\ar@{.}[rr]&&*=<1pt>{} }}
      &&
         \stackrel{\xymatrix@C=5pt@R=5pt{&&\ar@{.}[d]\\
                                         && \snode \ar@{-}[dl]\ar@{-}[dr]\\
                                         &*=<3pt>{\scriptstyle \ostar}\ar@{.}[dl]\ar@{.}[dr]&&\sCARRE&&&\\
                                         *=<1pt>{}\ar@{.}[rr]&&*=<1pt>{}&}}{~}
      \\
   \end{array}
  \end{displaymath}
  \caption[Foata-Zeilberger isomorphism]{Foata and Zeilberger isomorphism in pictures}
  \label{fig:FandZ}
\end{figure}
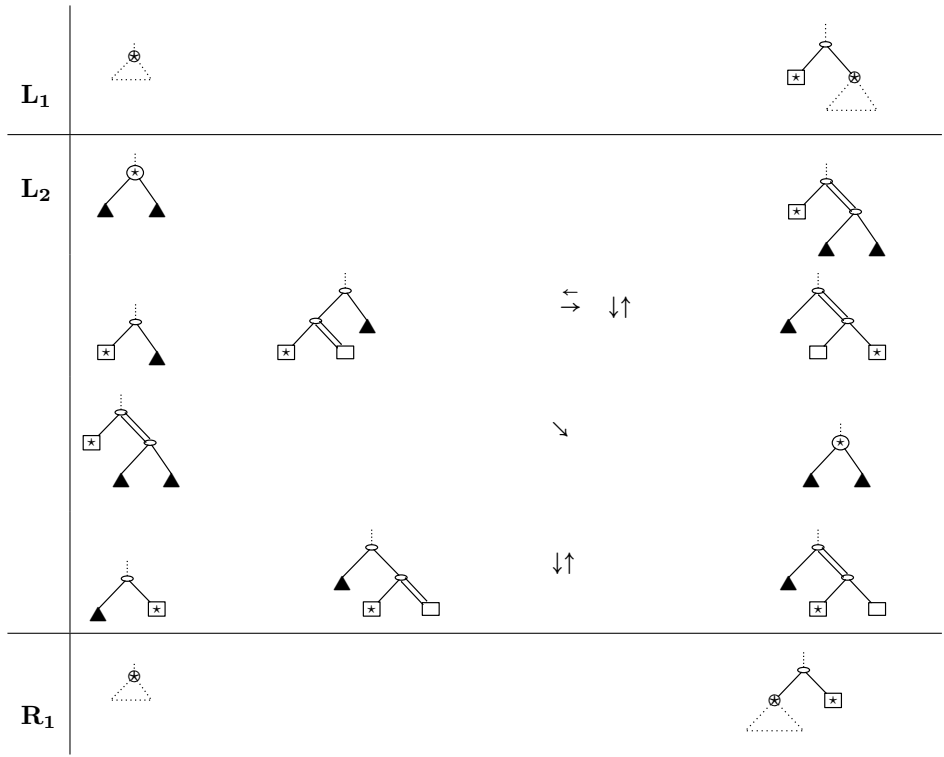

\begin{figure}[!]
  \centering
 \ifMint
  \begin{minted}{haskell}
type Gen = State StdGen

rst :: Int -> Int -> Gen (Vector (Int,Bool))
rst seed 0 = do put (mkStdGen seed) 
                return (initV // [(0,(0,False))])
rst seed n =
  let g = rst seed (n-1)
      body generator =
        do v <- g
           let (rand, newGenerator) = randomR (0::Double,1) generator
               x = floor (rand * fromIntegral (6*n-4))
               -- x is a random value between 0 and 6n - 4
               k = x `div` 3
           put newGenerator
           case x `mod` 3 of
             0 {-L1-} -> return (v // [(k,(2*n-1,False)),
                                       (2*n-1,(2*n,False)),
                                       (2*n,v!k)]) -- L1 
             1 {-L2-} -> case odd(fst (v!k)) of
               True {- not a leaf-}-> return (v// [(k,(2*n-1,False)),
                                                   (2*n-1,(2*n,False)),
                                                   (2*n,(fst(v!k),True))])
               False -> case odd k  of 
                  True {- the leaf is on the left -} -> case snd (v ! (k+1)) of
                    False {- the other link is black-} ->
                      return (v//[(k,v!(k+1)),
                                  (k+1,(2*n-1,True)),
                                  (2*n-1,v!k),(2*n,(2*n,False))])
        {-Failure-} True {- the other link is white -} -> body newGenerator
                  False {- the leaf is on the right -} ->
                    return (v//[(k,(2*n-1,True)),
                                (2*n-1,v!k),
                                (2*n,(2*n,False))])
             2 {-R1-} ->  return (v // [(k,(2*n-1,False)),
                                        (2*n-1,v!k),
                                        (2*n,(2*n,False))])
  in do generator <- get
        body generator
\end{minted}
\else
\begin{lstlisting}
type Gen = State StdGen

rst :: Int -> Int -> Gen (Vector (Int,Bool))
rst seed 0 = do put (mkStdGen seed) 
                return (initV // [(0,(0,False))])
rst seed n =
  let g = rst seed (n-1)
      body generator =
        do v <- g
           let (rand, newGenerator) = randomR (0::Double,1) generator
               x = floor (rand * fromIntegral (6*n-4))
               -- x is a random value between 0 and 6n - 4
               k = x `div` 3
           put newGenerator
           case x `mod` 3 of
             0 {-L1-} -> return (v // [(k,(2*n-1,False)),
                                       (2*n-1,(2*n,False)),
                                       (2*n,v!k)]) -- L1 
             1 {-L2-} -> case odd(fst (v!k)) of
               True {- not a leaf-}-> return (v// [(k,(2*n-1,False)),
                                                   (2*n-1,(2*n,False)),
                                                   (2*n,(fst(v!k),True))])
               False -> case odd k  of 
                  True {- the leaf is on the left -} -> case snd (v ! (k+1)) of
                    False {- the other link is black-} ->
                      return (v//[(k,v!(k+1)),
                                  (k+1,(2*n-1,True)),
                                  (2*n-1,v!k),(2*n,(2*n,False))])
        {-Failure-} True {- the other link is white -} -> body newGenerator
                  False {- the leaf is on the right -} ->
                    return (v//[(k,(2*n-1,True)),
                                (2*n-1,v!k),
                                (2*n,(2*n,False))])
             2 {-R1-} ->  return (v // [(k,(2*n-1,False)),
                                        (2*n-1,v!k),
                                        (2*n,(2*n,False))])
  in do generator <- get
        body generator
\end{lstlisting}
\fi
\caption[Schröder trees generation]{\textsf{Haskell} program for random generation of Schröder trees.}
\par\noindent\rule{\textwidth}{0.4pt}
  \label{fig:SchrProg}
\end{figure}

\section{A concrete algorithm for random generation of Schröder trees}
\label{sec:impSchroeder}

The program of Figure~\ref{fig:SchrProg} presents the algorithm for random
generation of Schröder trees.  Like in previous cases one uses a vector (an array) of
size $2n+1$. But in addition to the indices for the next nodes, one adds a
boolean.  This boolean says that the right link that starts from the node
corresponding to this index is white.  Hence each cell of the array
contains a pair $(k,b$) where $k$ is an index and $b$ is a boolean.
According to the constraints induced on the Schröder trees, the pair
corresponding to a boolean $True$ has, as a first component, an odd number,
since this first component corresponds to a link to a node.  Since the
node is a right child, it is located at an even index.  If these
constraints are not fulfilled, the second component must be a $False$.
Therefore in the program when she adds a pair $(k,True)$ at a
position $m$, she has to check that $k$ is odd and $m$ is even.  In another
hand, if she writes $(m,(k,True))$ for such an operation, she
has to check that it is of the form $(2p,(2q+1,True))$.  This constraints
is an invariant of the program.  The case when Foata and Zeilberger
produce a Schröder tree of size $n-2$ corresponds in my algorithm to a
``failure'', that is a ``retry'': the subprogram \textsf{body} is called
with a new random generator.

There are six cases.  Assume one draws a number $x$ between $0$ and $6n-4$
and let us call $k$ the number $x \div 3$. The value $x \mod 3$
discriminates among $L_1$, $L_2$ and $R_2$: $0$ for $L_1$, $1$ for $L_2$
and $2$ for $R_1$.  The cases $L_1$ and $R_1$ are easy.  Assume that
at the $k^{th}$ index the array contains $(h,b$).
\begin{description}
\item[$L_1$] corresponds in the code to
  \begin{center}
    \begin{tt}
    (v // [(k,(2*n-1,False)),(2*n-1,(2*n,False)),(2*n,v!k)]
  \end{tt}
\end{center}
The links are black hence the booleans are $False$.
\item[$L_2$ and $h$ is odd.] This means that the mark is on a node. 
This corresponds to the code:
  \begin{center}
    \begin{tt}
      v// [(k,(2*n-1,False)),(2*n-1,(2*n,False)),(2*n,(fst(v!k),True))]
    \end{tt}
  \end{center}
  Clearly, \texttt{2*n,(fst(v!k),True)} fulfills the constraints
  $(2p,(2q+1,True))$ since \texttt{fst(v!k)}, which we called $h$, is odd.
\item[$L_2$ and $h$ is even and $k$ is odd and] \textbf{the second component of \texttt{v!(k+1)} is \emph{False}:}
  This means that the mark is
  on a leaf ($h$ is even) which is a left child ($k$ is odd) and the link that goes to the sibling leaf is black.  This
  corresponds to the code:
  \begin{center}
    \begin{tt}
      v//[(k,v!(k+1)),(k+1,(2*n-1,True)),(2*n-1,v!k),(2*n,(2*n,False))]
    \end{tt}
  \end{center}
  The reader may check that the code corresponds to the picture of
  Figure~\ref{fig:FandZ}, line $2$.  \texttt{(k+1,(2*n-1,True))} fulfills
  the constraints $(2p,(2q+1,True))$ since $k+1$ is even and $2n-1$ is
  odd.
\item[$L_2$ and $h$ is even and $k$ is odd and the other link is white and] \textbf{the second component of \texttt{v!(k+1)} is \emph{True}:}
  This corresponds to a \emph{failure}, and the program loops with a new random
  generator.
\item[$L_2$ and $h$ is even and $k$ is even:]
  This means that the mark is a leaf ($h$ is even) which is a right child.   This corresponds to the code:
    \begin{center}
    \begin{tt}
v//[(k,(2*n-1,True)),(2*n-1,v!k),(2*n,(2*n,False))]
    \end{tt}
  \end{center}
  \texttt{(k,(2*n-1,True))} fulfills the constraints $(2p,(2q+1,True))$
  since $k$ is even and $2n-1$ is odd.
\item[$R_2$] corresponds in the code to
  \begin{center}
    \begin{tt}
      v // [(k,(2*n-1,False)),(2*n-1,v!k),(2*n,(2*n,False))
    \end{tt}
  \end{center}
The links are black hence the booleans are $False$.
\end{description}
One may notice that in the array, only values at odd indices require to
carry a boolean, and this boolean is required only when the first value of
the pair is odd.  This suggests a better data structure which may save little
space, but I did not implement it.

\subsection{Ratio of failures and average linear complexity}
\label{sec:ratFail}


The loop in \textsf{rst} can be sketched by the following program
\textsf{P}:
\ifMint
\begin{minted}{python}
while True:
    r = random.randint(0,2)
    if r == 0:
      break
    elif r == 1:
      continue
    else:
      break
\end{minted}
\else
\begin{lstlisting}
while True:
    r = random.randint(0,2)
    if r == 0:
      break
    elif r == 1:
      continue
    else:
      break
\end{lstlisting}
\fi
despite in \textsf{rst}, the loop happens  less often. In
\textsf{P}, one breaks with probability $\frac{2}{3}$, in the first run of
the loop, with probability $\frac{2}{3^2}$ in the second run of the loop,
with probability $\frac{2}{3^3}$ in the third run of the loop, etc.
Therefore if the cost of executing the statements of the loop is $c$, the total average cost of the  loop is:
\begin{eqnarray*}
  c\, \sum_{i=1}^{\infty}\frac{2}{3^i} &=& c.
\end{eqnarray*}

\subsection{Benchmarks}
\label{sec:bencSchr}

Benchmarks are done on a \textsf{Python} program, with a \textsf{while}
loop, which mimics the \textsf{Haskell} program. See
Figure~\ref{fig:BencSch}.  A third column yields the ratio of number of
fails over the size of the produced tree.  Computations are run on a laptop
with $4~Gb$ of memory.
\begin{figure}[!]
  \centering
  \begin{displaymath}
    \begin{array}{|r|r|l|}
      \hline\hline
      \mathbf{size}& \mathbf{time}&\mathbf{ratio}\\\hline\hline
      1\,000 & ~0.012s&0.024\\
      \hline
      5\,000 & ~0.031s&0.0288\\
      \hline
      10\,000& ~0.064s&0.025\\
      \hline
      50\,000 & ~0.200s&0.0269\\
      \hline
      100\,000& ~0.290s&0.02707\\
      \hline
      500\,000 & ~1.295s&0.027762\\
      \hline
      1\,000\,000&  ~3.065s&0.027883\\
      \hline
      5\,000\,000 & 15.183s&0.0276378\\
      \hline
      10\,000\,000 &  30.738s& 0.0275827\\
      \hline
    \end{array}
  \end{displaymath}
  \caption{Benchmarks for Schröder trees}
  \label{fig:BencSch}
\end{figure}

\section{Related Works}
\label{sec:relWork}

\subsection*{On Motzkin trees}
\label{sec:onM}

Our work was inspired by Jean-Luc Rémy's
algorithm~\cite{DBLP:journals/ita/Remy85}.  Laurent
Alonso~\cite{alonso94:_unifor_gener_motzk_word} proposed an algorithm for
generating uniformly Motzkin trees.  His method consists in generating the
number~$k$ of binary nodes with the correct probability law; he uses then
standard techniques to generate a unary-binary tree with $k$ binary nodes
among $n$ nodes. The number of trees with $k$ binary nodes is
over\-/approximated by values that follow a binomial distribution:
choosing $k$ is therefore done using random generations for a binomial law
and rejections. Therefore his algorithm is \emph{linear on the average},
with possible but extremely rare long sequences of rejection.  Dominique
Gouyou-Beauchamps and Cyril Nicaud~\cite{gouyou-beauchamps10} propose a
random generation for color Motzkin trees which is linear on the average
and Srečko Brlek et al.~\cite{DBLP:journals/acta/BrlekPR06} propose an
extension of Alonso's algorithm to generalized versions of Motzkin trees.

Axel Bacher, Olivier Bodini and Alice
Jacquot~\cite{DBLP:journals/corr/BacherBJ14,jacquot14:these,DBLP:journals/tcs/BacherBJ17}
propose an algorithm with similar ideas. Especially their Figure~2 shares
similarity with our Figure~\ref{fig:sndcontrib}. There ``operations''
$G_3$, $G_4$ and $G_5$ are connected with our cases \textbf{RL},
\textbf{LR} and \textbf{LL} respectively.  \rouge{\one} corresponds to
\raisebox{6pt}{$\xymatrix@C=2pt@R=2pt{&\snode\ar@{-}[dl]\ar@{-}[dr]\\\scarre
    &&\sCARRE}$} and \bl{\onw}~corresponds to
\raisebox{6pt}{$\xymatrix@C=2pt@R=2pt{&\snode\ar@{-}[dl]\ar@{-}[dr]\\\sCARRE
    &&\scarre}$}.  Like Alonso's, the algorithm they propose has a linear
expected complexity, due to failures similar to those of our Schröder tree
generation algorithm.  Let us also mention generic Boltzmann's samplers
with exact-size which apply among others to Motzkin
trees~\cite{DBLP:conf/analco/BendkowskiBD18,DBLP:journals/corr/Lescanne14}
and generic
algorithms~\cite{DBLP:journals/siamcomp/Devroye12,https://doi.org/10.48550/arxiv.2110.11472}
with linear expected complexity.

Denise and Zimmermann~\cite{DBLP:journals/tcs/DeniseZ99} discuss what can
be done on floating-point arithmetic when generating random
structures. The authors focus on decomposable labeled
structures~\cite{DBLP:journals/tcs/FlajoletZC94} and address the problem
of choice (which I call oracle), with a specific section on Motzkin trees.

\subsection*{On Schröder trees}
\label{sec:onS}

After studying the random generation of Motzkin trees and reading Foata
and Zeilberger paper, I started the implementation of a random generation
of Schröder trees, which turns out to be of expected linear
complexity. Actually Laurent Alonso, René Schott and Jean-Luc
Rémy~\cite{DBLP:journals/ipl/AlonsoRS97} proposed another linear algorithm
for random generation of Schröder trees, on the same principle as Alonso's
algorithm for the generation of Motzkin trees.  Like this quasi-linear
algorithm, it proceeds in two steps: first, it chooses randomly the number
$k$ of leaves with an adequate probability and a rejection technique,
second, it generates a random Schröder tree with $k$ leaves. In
comparison, my algorithm is direct. I deal only with Schröder trees, not
with Schröder trees with $k$ leaves.  I am very closed to Rémy's algorithm
and to my algorithm for random generation of Motzkin trees.

\section{Conclusion}
\label{sec:concl}
Generating Motzkin trees and Schröder trees has many potential
applications~\cite{donaghey77:_motzk,DBLP:journals/dm/BarcucciLPP00,DBLP:journals/ipl/AlonsoRS97}.
My algorithm for generation of Motzkin trees has a simple code and is
linear and my algorithm for Schröder trees is direct, which means it deals
only with Schröder trees.  Among the possible extensions of my method
which could be explored, there is the generation of extended versions of
Motzkin structures like Motzkin trees with colored
leaves~\cite{gouyou-beauchamps10} or Motzkin paths with $k$~long
straights~\cite{DBLP:journals/acta/BrlekPR06}. $k=1$ corresponds to
Motzkin paths and $k=2$ to Schröder paths.  On another hand, Bracucci et
al.~\cite{DBLP:journals/dm/BarcucciLPP00} study a family of sets of
permutations: $\mathcal{M}_1$, $\mathcal{M}_2$, ...,
$\mathcal{M}_{\infty}$, in which $\mathcal{M}_1$ is for Motzkin
permutations (that are Motzkin trees up to a bijection) and
$\mathcal{M}_{\infty}$ is for Catalan permutations (that are binary trees
up to a bijection) an interpolation of our method seems doable.

Deriving linear algorithms for the generation of random objects applies to
other structures. Indeed the production of a holonomic function can be
mechanized by the software GFUN~\cite{DBLP:journals/toms/SalvyZ94}.
However, for my purpose, this method has limitations, since the size of the
equation can blow up as shown by Figure~3
of~\cite{DBLP:journals/lmcs/BendkowskiL19} for instance.

\acknowledgments{I thank Laurent Alonso, Maciej Bendkowski, Alain Giorgetti and Jean-Luc Rémy
for interesting discussions and suggestions.}

\bibliographystyle{plainurl} 

\end{document}